\documentclass[useAMS,usenatbib]{mn2e}
\usepackage{savesym}
\usepackage{txfonts}
\savesymbol{iint}
\savesymbol{iiint}
\savesymbol{iiiint}
\savesymbol{idotsint}
\usepackage{graphicx}
\usepackage{amssymb}
\usepackage[below]{placeins}
\usepackage{txfonts}
\usepackage{natbib}
\usepackage{amsmath}
\usepackage{float}
\restoresymbol{TXF}{iint}
\restoresymbol{TXF}{iiint}
\restoresymbol{TXF}{iiiint}
\restoresymbol{TXF}{idotsint}
\usepackage{multirow}
\usepackage{array}
\bibpunct{(}{)}{;}{a}{}{,}

\def \xmm{{\emph{XMM-Newton}}}

\def \suzaku{{\emph{Suzaku}}}

\DeclareMathOperator{\arctanh}{arctanh}

\def\spose#1{\hbox to 0pt{#1\hss}}
\def\approxlt{\mathrel{\spose{\lower 3pt\hbox{$\sim$}}
        \raise 2.0pt\hbox{$<$}}}
\def\approxgt{\mathrel{\spose{\lower 3pt\hbox{$\sim$}}
        \raise 2.0pt\hbox{$>$}}}
\def\approxpropto{\mathrel{\spose{\lower 3pt\hbox{$\sim$}}
        \raise 2.0pt\hbox{$\propto$}}}
\mathchardef\twiddle="2218

\def\multleft#1{\hbox to size{\vbox {\halign {\lft{##}\cr #1}}\hfill}\par}
\def\multright#1{\hbox to size{\vbox {\halign {\rt{##}\cr #1}}\hfill}\par}

\def\today{\ifcase\month\or January\or February\or March\or April\or May\or
      June\or July\or August\or September\or October\or November\or December\fi
      \space\number\day, \number\year}
\def\<{\thinspace}

\def\ga{{\rm\thinspace gauss}}

\def\Msun{\hbox{$\rm\thinspace M_{\odot}$}}

\def\cha{{\it Chandra}}

\def\xmm{{\it XMM-Newton}}

\makeatletter
\newcommand{\thickhline}{%
    \noalign {\ifnum 0=`}\fi \hrule height 1.2pt
    \futurelet \reserved@a \@xhline
}
\newcolumntype{"}{@{\hskip\tabcolsep\vrule width 1pt\hskip\tabcolsep}}
\makeatother

\bibpunct{(}{)}{;}{a}{}{,}

\title[Dark Matter X-Ray Emission Lines in Galaxy Clusters]
{A \emph{Suzaku} Search for Dark Matter Emission Lines in the X-ray Brightest Galaxy Clusters}

\author[O. Urban et al.]{O. Urban$^{1,2,3}$\thanks{E-mail: ondrej@stanford.edu},
N. Werner$^{1,2}$, S. W. Allen$^{1,2,3}$, A. Simionescu$^{4}$,\newauthor
J. S. Kaastra$^{5}$, L. E. Strigari$^{6}$\\
$^{1}$Kavli Institute for Particle Astrophysics and Cosmology, Stanford University, 452 Lomita Mall, Stanford, CA 94305-4085, USA\\
$^{2}$Department of Physics, Stanford University, 382 Via Pueblo Mall, Stanford, CA 94305-4060, USA\\
$^{3}$SLAC National Accelerator Laboratory, 2575 Sand Hill Road, Menlo Park, CA 94025, USA\\
$^{4}$Institute of Space and Astronautical Science (ISAS), JAXA, 3-1-1 Yoshinodai, Sagamihara, Kanagawa, 252-5210 Japan\\
$^{5}$SRON Netherlands Institute for Space Research, Sorbonnelaan 2, 3584 CA Utrecht, The Netherlands\\
$^{6}$Mitchell Institute for Fundamental Physics and Astronomy, Texas A \& M University, College Station, TX 77843-4242\\
}

\begin{document}
\maketitle

\begin{abstract}

In light of recent claims of the discovery of an unidentified emission line
at $\sim$3.55~keV in stacked \xmm\ spectra for galaxy clusters, as well
as \xmm\ and \cha\ spectra for the Milky Way and M31, and the
possible association of this line with a decaying dark matter (DM - possibly
sterile neutrino) origin, we search for the presence of unidentified
emission lines in deep \suzaku\ X-ray spectra for the central regions of the
four X-ray brightest galaxy clusters: Perseus, Coma, Virgo and Ophiuchus. 
We employ an optimized energy range for our analysis ($3.2-5.3$keV) that is
relatively free of instrumental features, and a baseline plasma emission
model that incorporates the abundances of elements with the strongest
expected emission lines at these energies (S, Ar, Ca) as free parameters. 
For the Perseus Cluster core, employing our baseline plasma model, we find
evidence for an additional emission feature at an energy
$E=3.51^{+0.02}_{-0.01}$keV with a flux of
$2.87_{-0.38}^{+0.33}\times10^{-7}\text{ph}\,\text{s}^{-1}\text{cm}^{-2}\text{arcmin}^{-2}$. 
At slightly larger radii, we detect an emission line at $3.59\pm0.02$ keV
with a flux of
$4.8_{-1.4}^{+1.7}\times10^{-8}\text{ph}\,\text{s}^{-1}\text{cm}^{-2}\text{arcmin}^{-2}$. 
The energies and fluxes of these features are broadly consistent with
previous claims, although the radial variation of the line strength appears
in tension with standard dark matter model predictions.  Assuming a decaying
DM origin for the Perseus emission features allows us to predict the
energies and detected line fluxes for the other clusters in our sample. 
Critically, we do not detect an emission feature at the predicted energy and
line flux in the Coma, Virgo and Ophiuchus clusters.  The formal 99.5~per
cent upper limits on the strengths of an emission line in each cluster are below the
decaying DM model predictions, scaling from the Perseus Cluster center,
apparently ruling the model out.  In the light of these results, which
disfavor a decaying DM interpretation, we search for other explanations for
the $\sim 3.55$keV emission feature in Perseus.  Refitting the spectra with
a more complex plasma model that allows the abundances of additional
elements (Cl, K, Ti and V) to be free parameters removes the need for any
unidentified line in the $3.5-3.6$~keV region.  Our results suggest that
systematic effects associated with modeling the complex projected spectra
for the Perseus Cluster core, details of the assumed ionization balance, and
errors in the predicted emissivities of individual spectral lines may in
part be responsible for the $\sim 3.5$~keV feature.  The upcoming
\emph{Astro-H} satellite will allow us to explore the origins of this
feature more robustly.

\end{abstract}

\begin{keywords}
dark matter, line: identification, neutrinos
\end{keywords}

\section{Introduction}
\label{sect:intro}

Recent claims of the discovery of a previously unknown X-ray emission line
at $\sim3.5$~keV in the spectra of galaxy clusters \citep[hereafter
Bu14]{bulbul2014a} and nearby galaxies \citep[hereafter Bo14]{boyarsky2014a}
have sparked a lively discussion in the community.  These authors put
forward the possibility that the line originates from the decay of
$\sim$7~keV sterile neutrino dark matter (DM) particles in the halos of the
observed structures.

Bu14 found a weak unidentified line at $E=\left(3.55-3.57\right)\pm0.03$~keV
at a $>4\sigma$ significance using stacked \xmm\ observations of 73~low
redshift ($0.01<z<0.35$) clusters of galaxies.  The line was also seen in
subsets of their sample.  The authors predicted the intensity of nearby weak
elemental lines based on the strong lines in their fitting band, including
systematic uncertainties, and claimed that the unidentified line cannot have
an astrophysical origin.  Using \cha\ observations, the authors also
detected the line in the Perseus Cluster, although they made no detection in
the Virgo Cluster.

Bo14 reported a 4.4$\sigma$ detection of an unidentified line at
$3.518_{-0.022}^{+0.019}$~keV in combined \xmm\ observations of~M31 and
the Perseus Cluster.  For the Perseus data, the authors measured the surface
brightness profile of the feature, finding that it has a shallower slope
than predicted by a $\beta$-model (which approximately describes the surface
brightness of the intracluster medium, hereafter ICM).  They claim that the
radial variation of the line is consistent with a projected Navarro, Frenk
\& White \citep[NFW,][]{navarro1996} profile.

Recently, \citet{boyarsky2014b} also claimed a detection of a line at
$3.539\pm0.011$~keV in \xmm\ observations of the Galactic Center.  However,
in this case they could not rule out an astrophysical origin for the line.

Several other studies have been published recently that challenge the
presence of the line and/or the decaying DM interpretation: Using \cha\
observations of the Milky Way, \citet{riemer-sorensen2014} claim to exclude
the presence of a DM emission at $\sim$3.5~keV at 95\% confidence level. 
Instead, they argue, that the spectrum can be modeled using only continuum
emission and known elemental lines.  \citet{jeltema2014} analyzed stacked
\xmm\ observations of the Milky Way center.  They argue, that all spectral
features in the $\sim$3.5~keV region can be explained using standard
astrophysical lines, in particular those of K~\textsc{xviii}.  Evidence for
the line was also reevaluated for the Perseus observations analyzed in Bu14;
after including the previously unaccounted for Cl~\textsc{xvii} line at
3.51~keV, \citet{jeltema2014} found no evidence for an unidentified line in
this observation.  These authors also reanalyzed the M31 observations
of~Bo14, finding no significant evidence for the presence of the line. 
\citet{boyarsky2014c} have recently challenged this last claim, arguing that
the limited $3-4$~keV energy band used by \citet{jeltema2014} may lead to
biased results.  Specifically, they claim that the best-fit continuum model
obtained in this relatively narrow energy band overpredicts the flux
observed at higher energies, between $4-8$~keV.  It has also been suggested
by \citet{bulbul2014b} that the analysis of \citet{jeltema2014} may have
been compromised by the use of erroneous values for certain line
emissivities.

\citet{malyshev2014} used stacked \xmm\ observations of dwarf spheroidal
galaxies to exclude the Bu14 line at $3.3\sigma$.  Due to their large
mass-to-light ratio and the fact that their interstellar medium does not
produce thermal X-ray radiation, these objects provide an interesting tool
for such tests.  However, to conclusively exclude a DM line origin, the
authors call for additional observations that would increase the sensitivity
by a factor of $\sim$2.

Finally, \citet{anderson2014} analyzed observations of 81 \cha\ and 89 \xmm\
galaxies, respectively stacked in a way where each X-ray event is weighed
according to the assumed DM profile in its respective galaxy.  The authors
ruled out the presence of the purported DM decay line at $4.4\sigma$ and
$11.8\sigma$ confidence in the \cha\ and \xmm\ samples, respectively.

A brief review of the current status of the observational results and theoretical interpretations of the $\sim$3.5~keV feature, including implications for dark matter small-scale structure \citep{abazajian2014}, can be found in~\citet{iakubovskyi2014}.

In this paper, we present our own search for unidentified emission lines in
\suzaku\ observations of the four X-ray brightest galaxy clusters: Perseus,
Ophiuchus, Virgo and Coma \citep{edge1990} -- all of which have relatively
long exposures.  Additionally, unlike the previous works, we fit all spectra
in parallel, avoiding stacking and the complications it brings.  In
particular, systematic effects, due to the smearing of spectral features by
e.g.  gain variations, are reduced in our analysis.

Sect.~\ref{sect:obsanal} of this paper describes our data analysis and
modeling approach.  In Sect.~\ref{sect:results} we present the main results
from our analysis and in~Sect.~\ref{sect:discus} we discuss their
implications.  Our conclusions are summarized in~Sect.~\ref{sect:concl}. 
Throughout this paper, all errors are quoted at 68~per~cent level of
confidence and all line energies are quoted and plotted in the rest frame,
unless stated otherwise.

\section{Observations and Data Analysis}
\label{sect:obsanal}

\begin{table}
\centering
\caption{List of observations used in the analysis. The columns show,
respectively, the \suzaku\ observation ID, the date of the observation and
the clean exposure time.  The total exposure time for the Perseus Cluster is
740~ks.}
\label{tab:data}
\begin{tabular}{lcc}
\hline\hline
\textbf{Obs. ID}&\textbf{Obs. date}&\textbf{Exposure (ks)}\\
\hline
\multicolumn{3}{c}{Perseus}\\
\hline
800010010$^\dagger$&2006-02-01&44\\
101012010$^\dagger$&2006-08-29&46\\
101012020&2007-02-05&40\\
102011010&2007-08-15&36\\
102012010&2008-02-07&36\\
103004010&2008-08-13&41\\
103004020&2009-02-11&46\\
104018010&2009-08-26&35\\
104019010&2010-02-01&34\\
105009010&2010-08-09&30\\
105009020&2011-02-03&33\\
105010010&2010-08-10&25\\
105027010&2011-02-22&42\\
106005010&2011-07-27&34\\
106005020&2012-02-07&42\\
106006010&2011-07-26&33\\
107005010&2012-08-20&34\\
107005020&2013-02-11&36\\
108005010&2013-08-15&38\\
108005020&2014-02-05&34\\
\hline
\multicolumn{3}{c}{Coma}\\
\hline
801097010$^\dagger$&2006-05-31&164\\
\hline
\multicolumn{3}{c}{Virgo}\\
\hline
801038010&2006-11-29&90\\
\hline
\multicolumn{3}{c}{Ophiuchus}\\
\hline
802046010&2007-09-24&83\\
\hline\hline
\multicolumn{3}{l}{$^\dagger$observations with available XIS2 exposure}
\end{tabular}
\end{table}

The details of the \suzaku\ observations utilized in this study are shown
in~Tab.~\ref{tab:data}.  For each cluster, we analyzed data from all X-ray
Imaging Spectrometers (XIS 0, 1, 2 and 3) obtained in the normal clocking
mode.  XIS~2 was lost to a putative micrometeoroid hit on November~9, 2006
and therefore its data are available only for a limited fraction of our
sample.

There is only one observation for each of the Coma, Virgo and Ophiuchus
cluster cores.  The Perseus Cluster is a \suzaku\ calibration source and
as such it has been observed regularly since the launch of the satellite. 
We use all observations for Perseus obtained in the normal clocking mode
with no window option, with a total exposure time of 740~ks.

\subsection{Data Cleaning}

Initial cleaned event lists were obtained using the standard screening
criteria proposed by the XIS team\footnote{Arida, M., XIS Data Analysis,\\
http://heasarc.gsfc.nasa.gov/docs/suzaku/analysis/abc/node9.html}.  Given
the energy band used for the analysis ($3.2-5.3\,\text{keV}$, see
Sect.~\ref{subs:modeling}), we do not expect any significant influence from
the solar wind charge-exchange (SWCX) emission, whose lines are expected
below $1.5\text{keV}$ \citep{fujimoto2007}.  We filtered out times of low
geomagnetic cut-off rigidity ($\text{COR}\leq 6\,\text{GV}$).  For the XIS~1
data obtained after the reported charge injection level increase on June 1st
2011, we have excluded two adjacent rows on either side of the
charge-injected rows (the standard is to exclude one row on either side).

\subsection{Spectral Analysis}

For the Coma, Virgo and Ophiuchus clusters, we extracted spectra from the
full field of view of the XIS detectors, excluding the regions associated
with the detector edges.  In the case of the Perseus Cluster we opt to
define two extraction regions: a central, circular region with a radius of
6~arcmin (which we refer to as the `core' region), and the rest of the
detector, excluding the edges (which we term the `confining' region). 
Spectra from each observation and for each detector were treated separately. 
We rebinned the spectra to have at least one count per bin, employing the
extended C-statistic \citep{cash1979,arnaud1996} in the fitting.  We use
night Earth observations to create the instrumental background spectra.

For each spectrum, we constructed an individual response matrix with a
resolution of $16\,\text{eV}$, spanning the $2.0-6.5\,\text{keV}$ energy
band\footnote{These rebinned matrices require $\sim 50$~times less disk
space and significantly improve the speed of the analysis with respect to
the default choice of $2\,\text{eV}$ resolution and the full
$0.2-16.0\,\text{keV}$ \suzaku\ energy band, without compromising the
accuracy of the results}.  Ancillary response files were constructed for
each spectrum with the task \textsc{xissimarfgen}, at the resolution
appropriate for the response matrices.

\subsection{Spectral Modeling}
\label{subs:modeling}

Decay lines from decaying dark matter are expected to be relatively weak in
the X-ray spectra of clusters.  Therefore, we selected an energy band for
our analysis that retains as much signal as possible, while minimizing
sensitivity to the instrumental systematics.  In detail, at energies just
below $\sim 3.2\,\text{keV}$ the effective area of the \suzaku\ mirrors is
influenced by the presence of strong M-edges of gold \citep{tamura2006},
while above 5.3 keV the modeling becomes sensitive to instrumental lines
including the Mn-K$\alpha$ line at 5.9~keV \citep{tawa2008}.  For these
reasons, we limited our analysis to the 3.2-5.3keV band, which includes the
region around $\sim$3.5~keV highlighted in previous works.

We carried out our spectral modeling in \textsc{xspec} \citep[version
12.8]{arnaud1996}.  All spectra from a given extraction region were fitted
simultaneously.  We modeled the ICM emission as a single temperature plasma
in collisional ionization equilibrium using the absorbed \textsc{vapec} or
\textsc{vvapec}\footnote{The models differ only in the number of elemental
abundances they allow to fit for independently; 14 and 30 for \textsc{vapec}
and \textsc{vvapec}, respectively.} models \citep{smith2001}.  The
temperatures and normalizations of the individual spectra were left free and
independent, which mitigates the effects of small differences in calibration
between the individual detectors.  The Galactic absorption was set to the
average column along the line of sight inferred from the
Leiden/Argentine/Bonn Survey \citep{kalberla2005}.  For each cluster the
redshift was fixed to the value provided by the NASA/IPAC Extragalactic
Database (NED).  We adopted the solar abundance table of
\citet{feldman1992}.  In contrast with previous works, where the data were
modeled by a line-free continuum and individual gaussian line components,
our approach utilizes all lines of a given element simultaneously.

\textsc{AtomDB}~v.~2.0.2\footnote{http://www.atomdb.org/} \citep{foster2012}
shows that there are three heavy elements with lines with emissivities
higher than $5\times10^{-19}\,\text{photons}\,\text{cm}^3\text{s}^{-1}$ (for
the appropriate plasma temperatures of the clusters) in the observed energy
band: S, Ar and Ca.  The abundances of these elements were included as free
parameters in the fits.  The abundances of other elements were initially
fixed to 0.5~Solar, with the exception of H and He, whose abundances were set
to unity.

The cosmic X-ray background (CXB) due to the unresolved point sources has
been modeled with an absorbed power law (PL) component
\citep[e.g.,][]{deluca2004}.  Other standard components of the CXB, e.g. 
the Galactic halo emission \citep{kuntz2000} and the emission from the local
hot bubble \citep{sidher1996} do not significantly contribute to the
observed signal in the energy band of the fit.  The parameters for the CXB
power law component have been fixed to the published values of
\citet{urban2014} for Perseus and \citet{simionescu2013} for Coma, and are
shown in Tab.~\ref{tab:cxfb}\footnote{No scaling of the normalization was
required here.  The ancillary response files in all the mentioned analyses
were created using the standard assumption, utilized in the \suzaku\
analysis, of uniform emission from a circular source with a radius of
20~arcmin.}.  The values for Perseus and Coma are consistent with each other
within their systematic uncertainties.  Assuming uniformity of the power law
component across the sky, we used the Perseus values to describe the local
CXB of the Virgo and Ophiuchus clusters.

\begin{table}
\centering
\caption{CXB components for the individual clusters. The normalization of the power law is given in units of
$10^{-3}\text{photons}\times\frac1{20^2\pi}\text{keV}^{-1}\text{cm}^{-2}\text{s}^{-2}\text{arcmin}^{-2}$ at 1~keV.}
\label{tab:cxfb}
\begin{tabular}{l|cccc}
\hline\hline
&\textbf{Perseus}&\textbf{Coma}&\textbf{Virgo}&\textbf{Ophiuchus}\\
\hline
PL index&1.52&1.51&1.52&1.52\\
norm$_{\rm PL}$ &1.28&1.15&1.28&1.28\\
\hline\hline
\end{tabular}
\end{table}

\section{Results}
\label{sect:results}

\subsection{Modeling With No Additional Line Component}
\label{sect:noline}

\begin{table*}
\setlength{\extrarowheight}{4pt}
\centering
\caption{Best-fit parameters for the individual extraction regions using the
baseline plasma model, with no additional features (see
Sect.~\ref{subs:modeling}).  60 individual spectra were modeled for both the
core and the confining regions of the Perseus Cluster, with results are
shown in Fig.~\ref{fig:pers}.  Normalizations are in the units of $\int
n_{\rm e}n_{\rm
H}dV\times\frac{10^{-14}}{4\pi\left[D_A(1+z)\right]^2}\frac1{20^2\pi}\,\text{cm}^{-5}\text{arcmin}^{-2}$.}
\label{tab:results}
\begin{tabular}{cc|ccccc}
\hline\hline
&&\textbf{Perseus (core)}&\textbf{Perseus (confining)}&\textbf{Coma}&\textbf{Virgo}&\textbf{Ophiuchus}\\
\hline
\multirow{4}{*}{\textbf{kT (keV)}}&XIS0&\multicolumn{2}{c}{\multirow{4}{*}{see Fig.~\ref{fig:pers}}}&$8.07\pm0.17$&$2.61\pm0.04$&$9.66\pm0.25$\\
                          &XIS1&\multicolumn{2}{c}{}                                            &$8.00\pm0.17$&$2.61\pm0.04$&$9.80\pm0.25$\\
			  &XIS2&\multicolumn{2}{c}{}                                            &$8.38\pm0.17$&\multicolumn{2}{c}{N/A}\\
			  &XIS3&\multicolumn{2}{c}{}                                            &$8.60_{-0.18}^{+0.29}$&$2.59\pm0.04$&$9.66\pm0.25$\\
\hline
\multirow{3}{*}{\textbf{abundance (Z/Z$_\odot$)}}&S &$2.12\pm0.21$&$2.24\pm0.63$&$2.1\pm1.4$&$1.18\pm0.38$&$9.6\pm2.0$\\
                           &Ar&$0.43\pm0.03$&$0.36\pm0.09$&$0.58\pm0.20$&$0.42\pm0.06$&$-0.24\pm0.26$\\
			   &Ca&$0.65\pm0.01$&$0.44\pm0.04$&$0.38\pm0.09$&$0.85\pm0.04$&$0.68\pm0.11$\\
\hline
\multirow{4}{*}{\textbf{normalization}}&XIS0&\multicolumn{2}{c}{\multirow{4}{*}{see Fig.~\ref{fig:pers}}}&$0.65\pm0.01$&$1.18\pm0.02$&$1.63\pm0.02$\\
                               &XIS1&\multicolumn{2}{c}{}                                           &$0.67\pm0.01$&$1.23\pm0.03$&$1.68\pm0.02$\\
			       &XIS2&\multicolumn{2}{c}{}                                           &$0.65\pm0.01$&\multicolumn{2}{c}{N/A}\\
			       &XIS3&\multicolumn{2}{c}{}                                           &$0.64\pm0.01$&$1.28\pm0.03$&$1.68\pm0.02$\\
\hline
\multicolumn{2}{c|}{\textbf{C-statistic/d.o.f.}}&34281.27/34078&33612.86/33874&2402.42/2289&1742.82/1716&1694.60/1716\\
\multicolumn{2}{c|}{\textbf{$\chi^2$/d.o.f.}}   &34566.72/33808&28429.36/28653&2414.91/2289&1755.67/1716&1703.69/1716\\
\hline\hline
\end{tabular}
\end{table*}

\begin{figure}
\centering
\includegraphics[width=.95\columnwidth]{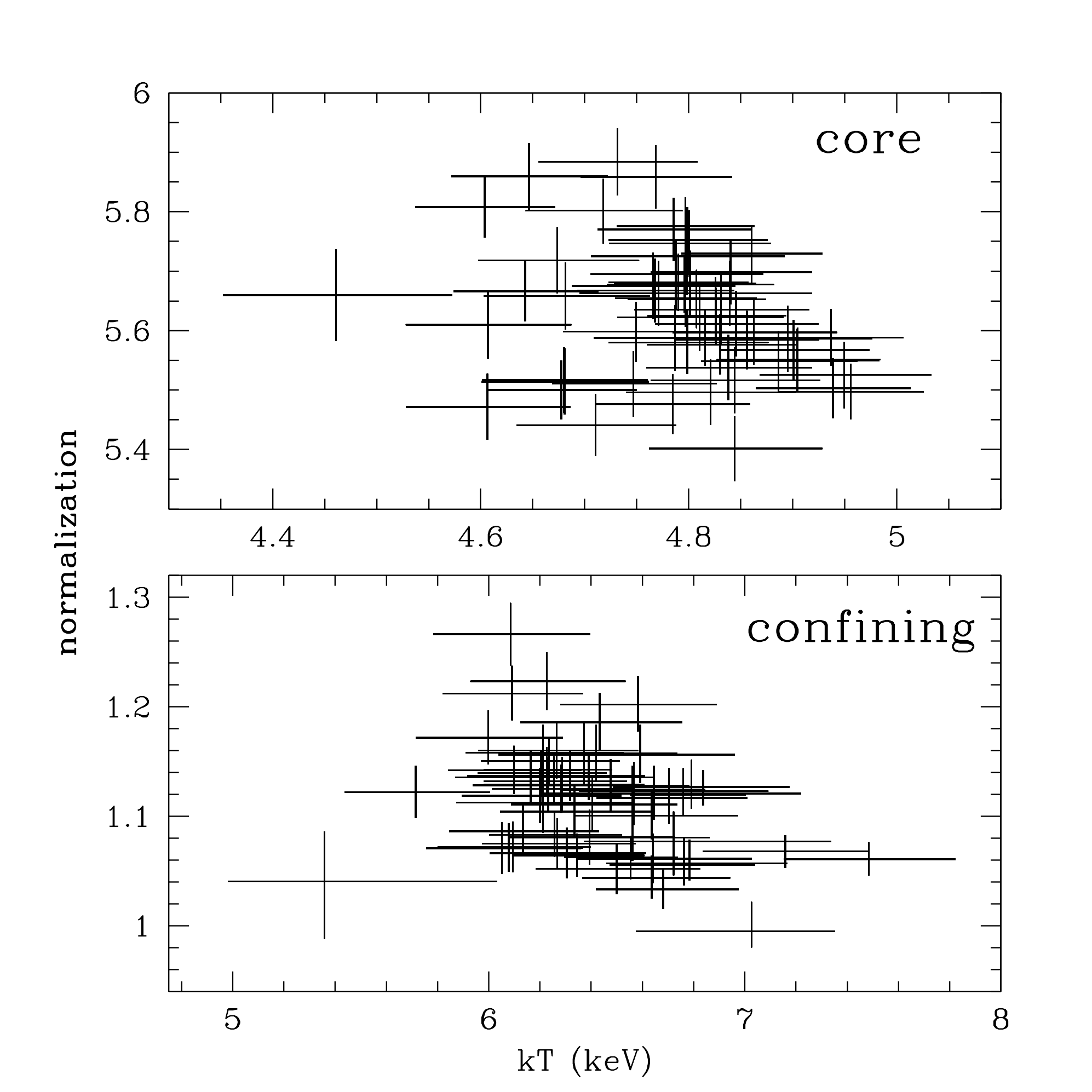}
\caption{Temperatures and normalizations for the Perseus Cluster measured in the different ObsIDs and with different XIS~detectors. Top panel shows the values for the core region ($r<6\,\text{arcmin}$) and the bottom panel for the
confining region ($r>6\,\text{arcmin}$). The normalizations are given in the units introduced in Tab.~\ref{tab:results}.}
\label{fig:pers}
\end{figure}

In the first step of the analysis, we applied our baseline plasma model,
including features due to S, Ar and Ca, with no additional lines.  The
results are summarized in Tab.~\ref{tab:results} and Fig.~\ref{fig:pers}. 
To estimate the goodness of fit for these baseline models, we rebinned the
spectra to have at least 50~counts per bin and determined the reduced
$\chi^2$, which are summarized in the bottom row of~Tab.~\ref{tab:results}. 
\footnote{In most cases, the rebinning did not have any effect, since the
Suzaku observations of bright cluster cores already contain at least
50~counts per bin.  This is demonstrated by the unchanged number of degrees
of freedom (d.o.f.) between the C-statistic and $\chi^2$ fits for the
longest individual observations (Coma, Virgo, Ophiuchus).} We obtained
broadly (though in some cases not formally) acceptable values of
$\chi^2$/d.o.f.  between 0.99 (the confining region of Perseus, and
Ophiuchus) and 1.06 (Coma), indicating that our baseline models describe the
data reasonably well.

The temperatures measured in the individual cluster regions with the
baseline models are in good general agreement with values found in
literature \citep[e.g.,][for Perseus, Coma, Virgo and Ophiuchus,
respectively]{fabian2011,sanders2013,million2010b,million2010a}.  Our
relatively narrow fitting band does not allow us to assess the
multitemperature structure present in the cluster cores.  We attempted to
fit the spectra with two temperature models, with the temperature of one of
the components fixed first at 90 and later at 50~per~cent of the value of
the other.  In neither case were we able to robustly constrain the
contribution of the individual components to the total emission. 
Multitemperature structure of the cluster cores has been studied extensively
in previous work on Perseus \citep[e.g.,][]{sanders2007} and Virgo
\citep{werner2010}.  It is clear, therefore, that our continuum model must
be biased to some degree, which must influence the predicted plasma model
line strengths.  This is discussed in more detail in~Sect.~\ref{subs:7abund}
and~\ref{subs:disc_3.55}.

Our initial fits with the baseline plasma model measured high, supersolar
abundances of S, particularly in Ophiuchus and Perseus.  In the former case,
the high S abundance appears to compensate in part for negative residuals at
the expected energies of some Ar~lines, since the lines from the ions of
both elements, in particular Ar~\textsc{xviii} and S~\textsc{xvi}, overlap
in a relatively narrow energy band (between $3.27-3.44~\text{keV}$).  The
high S~abundance found in Perseus could also be affected by other chemical
elements with weak lines.  This is discussed further
in~Sect.~\ref{subs:7abund}.

\subsection{Modeling With An Additional Line Component}
\label{subs:significance}

\begin{figure*}
\begin{minipage}{0.95\columnwidth}
\includegraphics[width=\textwidth]{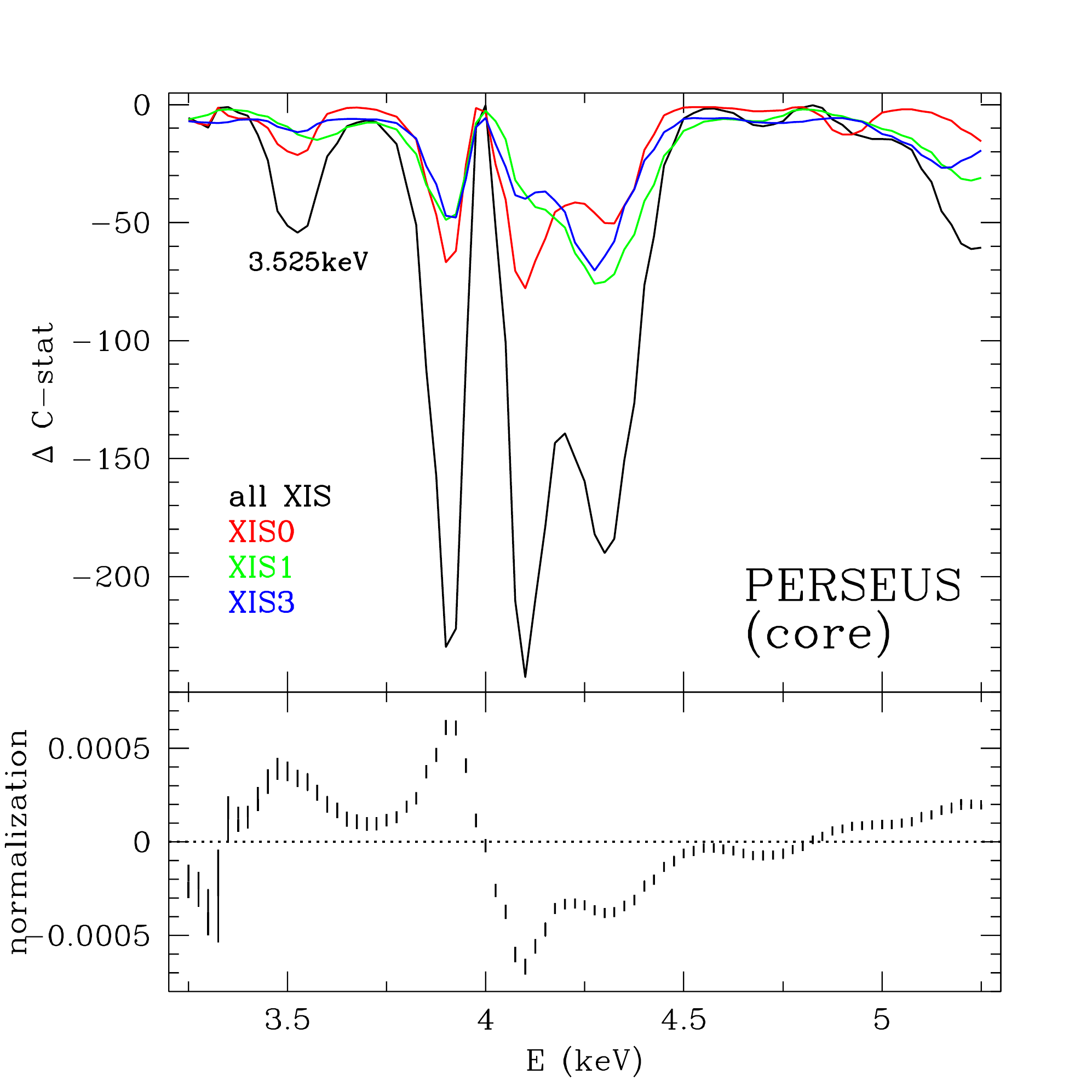}
\end{minipage}
\begin{minipage}{0.95\columnwidth}
\includegraphics[width=\textwidth]{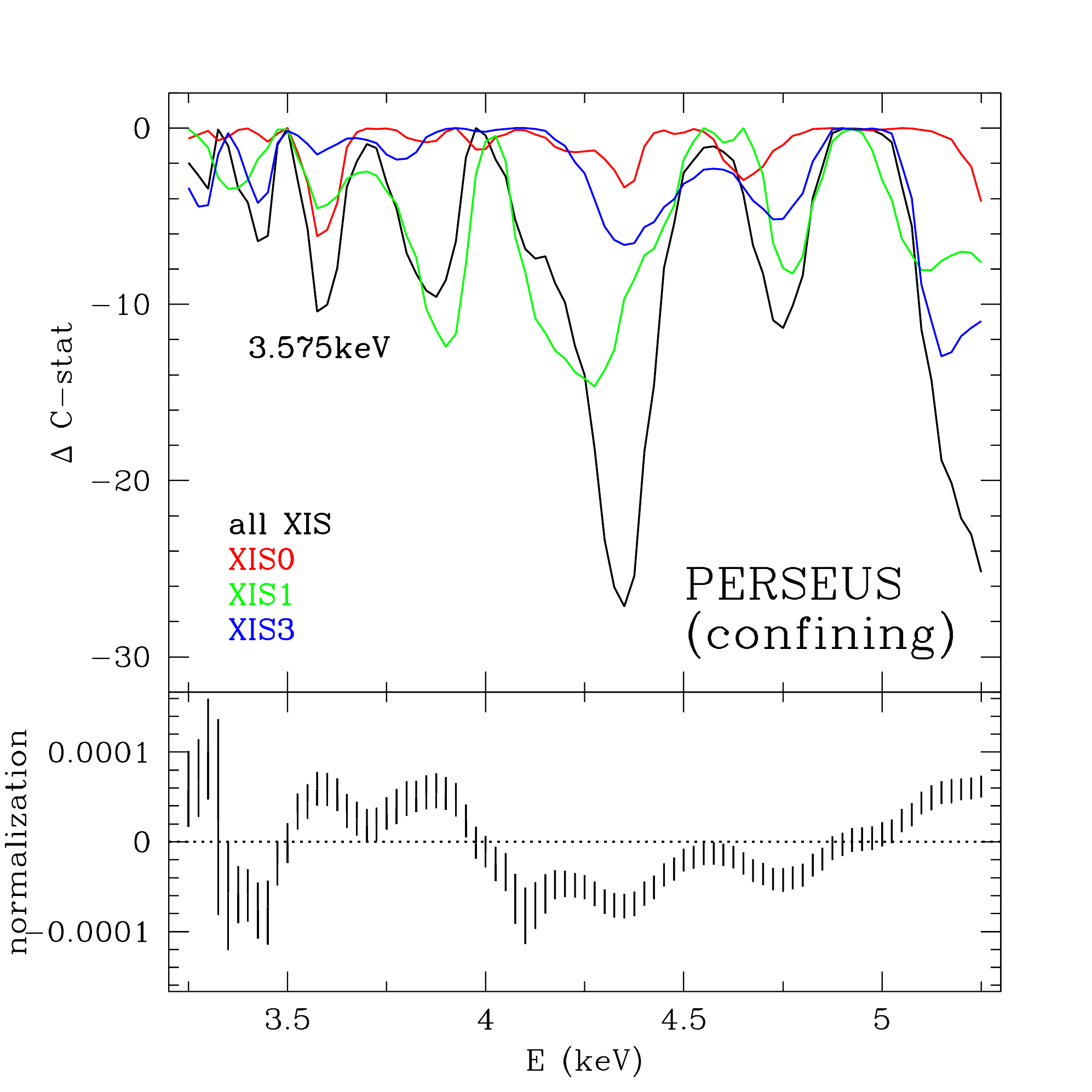}
\end{minipage}
\caption{Improvement of the fit for one additional degree of freedom (the
line normalization) as a function of the \emph{rest-frame} line energy for the core
(\emph{left}) and the confining (\emph{right}) regions of the Perseus
Cluster.  The top panels show the change in the C-statistic value between
the models with and without the additional line component.  Black lines show
the results when simultaneously using the data from all the detectors and
colored lines (red, green, orange, blue) using the data from the individual
detectors.  Bottom panels show the best-fit value for the line normalization
in units of
$\frac1{20^2\pi}\text{photons}\,\text{cm}^{-2}\text{s}^{-1}\text{arcmin}^{-2}$
using the data from all detectors.  Note the different scales of the
vertical axes.  Note also the presence of formal requirements for both
emission (positive residuals) and absorption (negative residuals) features
in the fits at different energies.}
\label{fig:significance}
\end{figure*}

\begin{figure*}
\begin{minipage}{0.95\columnwidth}
\includegraphics[width=\textwidth]{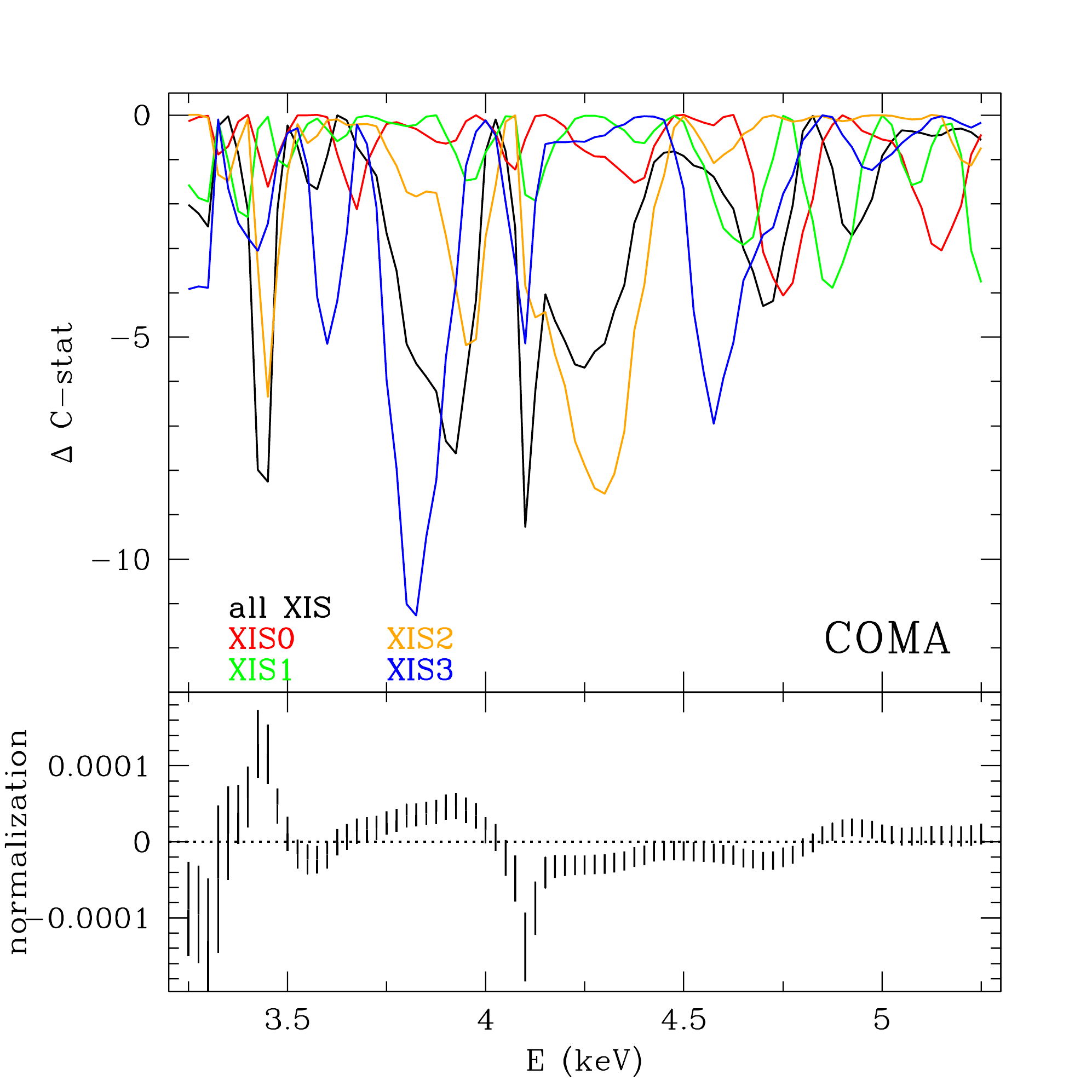}
\end{minipage}
\begin{minipage}{0.95\columnwidth}
\includegraphics[width=\textwidth]{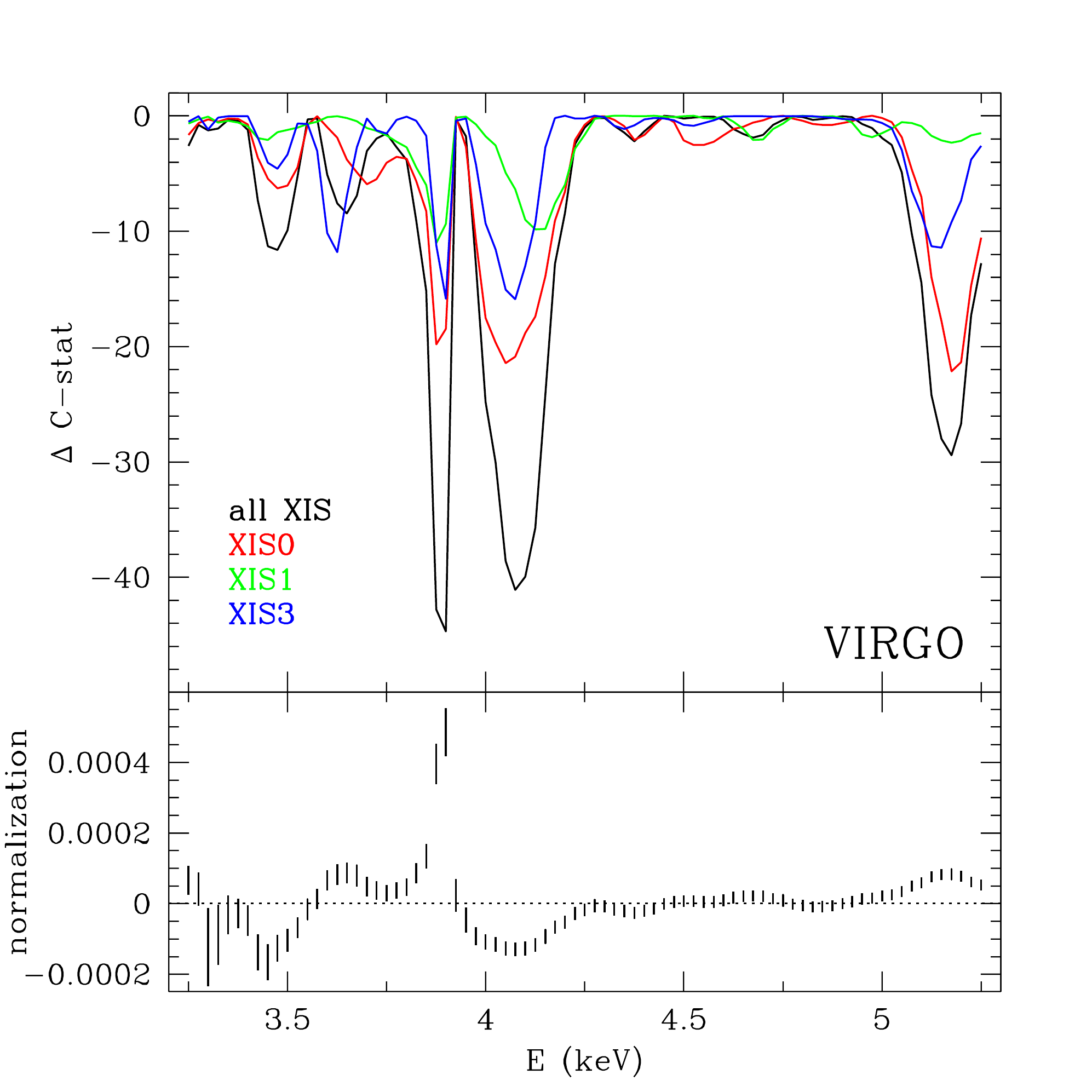}
\end{minipage}
\begin{minipage}{\textwidth}
\centering
\includegraphics[width=.475\textwidth]{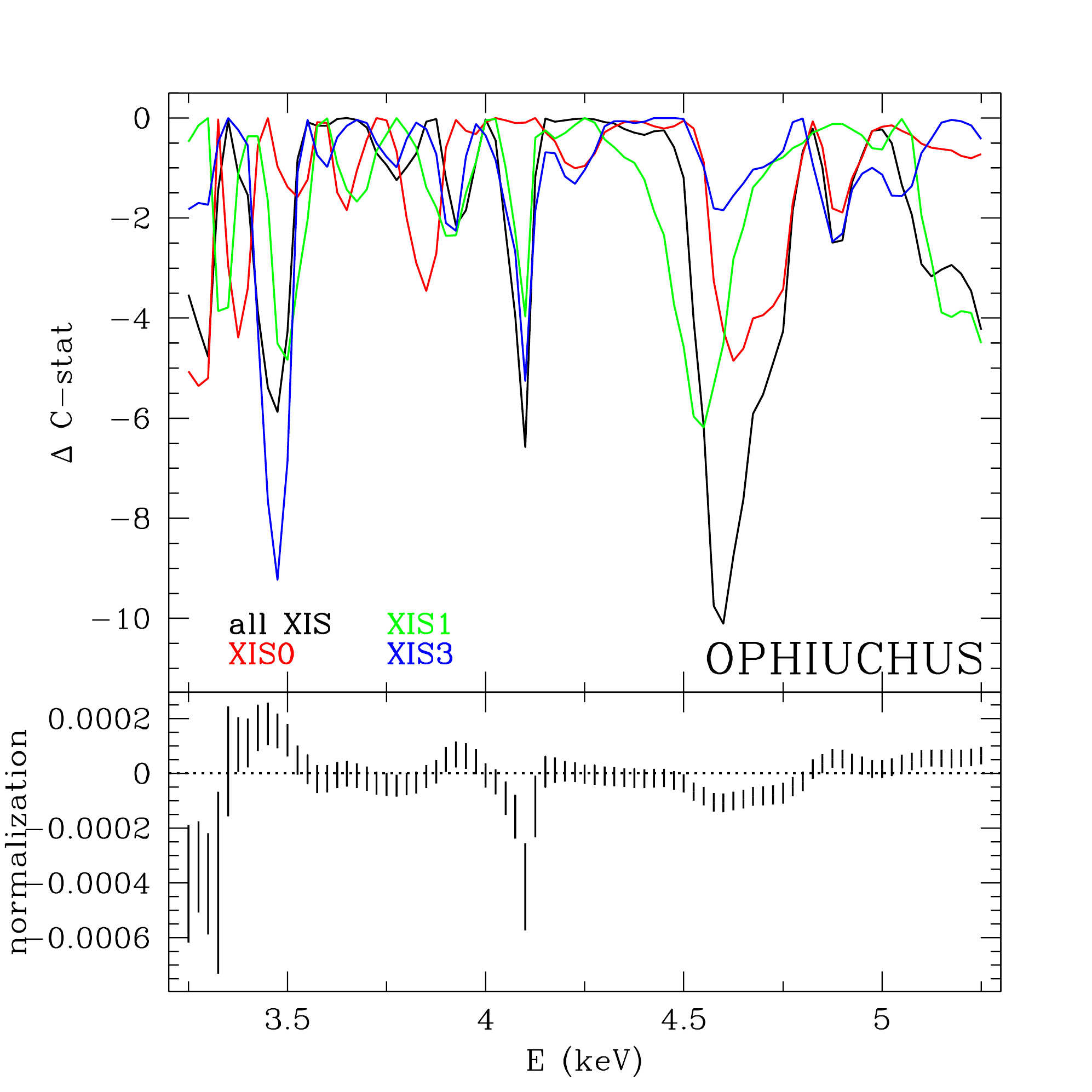}
\end{minipage}
\caption{As Fig.~\ref{fig:significance} for the Coma, Virgo and Ophiuchus Clusters.}
\label{fig:significance2}
\end{figure*}

To test for the presence of a spectral line in addition to the baseline
model within our energy band, we added a redshifted gaussian component into
the model and performed a sequence of fits for each cluster (including both
extraction regions of Perseus), successively fixing the energy of this line
component to values spanning the range $3.25-5.25~\,\text{keV}$ (in
intervals of 25~eV).  We allowed for both positive and negative line
normalizations, and measured the improvement ($\Delta C$) of the fit as a
function of the line energy.  In all cases we assume that the line has zero
intrinsic velocity width (meaning that the width of the line in the spectral
fit is set by the spectral resolution of the Suzaku detectors).

For each cluster, fits were performed on both a combined data set (fitting
the data for all detectors simultaneously) and the data for the separate XIS
detectors individually.  The results are shown in
Figs.~\ref{fig:significance} and \ref{fig:significance2}.

In the rest of this section, we focus our discussion on the presence of a
possible emission feature at $E\sim$3.55~keV, highlighted in previous works
(see~Sect.~\ref{sect:intro}).  A discussion of the other, and in some case
formally more significant features, can be found
in~Sect.~\ref{subs:disc_lines}.

\subsubsection{3.55~keV features in the Perseus Cluster}
\label{subs:3.55}

The most prominent improvement to the baseline model with the introduction
of an additional line at energies around 3.55keV is found in the data for
the Perseus Cluster core region, shown in the left panel of
Fig.~\ref{fig:significance}.  We measure $\Delta C=-54.82$, with respect to
the baseline model, with the introduction of a line at 3.525~keV. 
Performing a separate fit with the line energy included as a free parameter
but constrained to be in the range $3.4-3.6$~keV, we find
$E=3.510_{-0.008}^{+0.023}\text{keV}$ and a flux of
$2.87_{-0.38}^{+0.33}\times10^{-7}\text{ph}\,\text{s}^{-1}\text{cm}^{-2}\text{arcmin}^{-2}$
($\Delta C=-55.36$ for two additional d.o.f with respect to the baseline
model).

A less significant feature, at $E=3.575$~keV, is found in the confining
region of Perseus, plotted in the right panel of Fig.~\ref{fig:significance}
($\Delta C=-10.41$).  Including the line energy as a free parameter, we
measure a line energy $E=3.592_{-0.024}^{+0.021}~\text{keV}$ and flux
$4.78_{-1.41}^{+1.65}\times10^{-8}\text{ph}\,\text{s}^{-1}\text{cm}^{-2}\text{arcmin}^{-2}$
($\Delta C=-10.93$ for two additional d.o.f).

The energies of the lines in the core and confining regions are formally
separated by more than 3~$\sigma$.  Given that tests based on the observed
energy of the Fe-K~line at 6.7~keV indicate the gain to be stable within
$\sim$5~eV during the Perseus observations, this separation is significant. 
Nonetheless, the 82~eV difference is smaller than the resolution of the
\suzaku\ CCD detectors ($\sim$150~eV).

In the following argument, we adopt the assumption that these two lines have
the same origin.  In Sect.~\ref{subs:expectation} we show, that assuming a
DM~decay origin of these lines, and an NFW profile for the DM~density, we
would expect the dark matter lines from the core and confining regions of
Perseus to be comparable in strength, which is in tension with our results. 
Using the best-fit energies of the respective lines, they differ in fluxes
by a factor of 6.  This difference increases to a factor of $\sim$14, when
the energy of the line in the confining region is fixed to 3.51~keV (the
best fit value in the core).  On the other hand, when we fix the energy of
the line in the core to 3.592~keV (the best-fit energy from the confining
region) the difference in fluxes slightly decreases to a factor of $\sim$4.

\subsubsection{Absence of the 3.55~keV features from the Coma, Virgo and Ophiuchus clusters}
\label{subs:no3.55}

We found no comparably significant emission features with energies around
3.55~keV in the Suzaku data for the Coma, Virgo or Ophiuchus clusters, with
formal 95 per cent confidence upper limits on the strength of any line at
$E=3.55$ keV of
$2.65\times10^{-9}\text{ph}\,\text{s}^{-1}\text{cm}^{-2}\text{arcmin}^{-2}$,
$2\times10^{-8}\text{ph}\,\text{s}^{-1}\text{cm}^{-2}\text{arcmin}^{-2}$ and
$7.06\times10^{-8}\text{ph}\,\text{s}^{-1}\text{cm}^{-2}\text{arcmin}^{-2}$,
respectively.  By fitting the energies of the strongest astrophysical
emission lines in the Coma, Virgo and Ophiuchus clusters we determined, that
gain calibration changes can shift the energies of various spectral features
by at most 30~eV in either direction in energy space.  Therefore, it is
improbable that other features present in the spectra of the Coma, Virgo and
Ophiuchus clusters beyond this range (e.g.  the 3.45~keV emission feature
seen in the upper left panel of Fig.~\ref{fig:significance2} for the Coma
Cluster) has the same origin as the 3.55~keV features in the Perseus
Cluster.

\subsubsection{The 3.55~keV Line Strength Expectation}
\label{subs:expectation}

The energies and fluxes of the additional (w.r.t. the baseline plasma model;
see Sect.~\ref{subs:modeling}) $E\sim 3.55$keV emission features detected in
the Suzaku data for the Perseus Cluster core and confining region are
similar to those reported by~Bu14 and Bo14.

If we assume a dark matter origin for the Perseus emission features, we can
predict the energies and line fluxes with which these lines should appear in
the spectra for the other clusters in our sample, and test for consistency
with the measured upper limits.  Such a comparison provides a rigorous test
of the DM hypothesis.

Assuming a decaying dark matter (e.g. sterile neutrino) origin for the
Perseus emission features, the strengths of the lines in the other clusters,
relative to the Perseus core (and the confining region of Perseus), should
depend only on the projected masses of dark matter within the respective
spectral extraction regions, and the distances to the clusters.  The
three-dimensional density profiles of the dark matter halos hosting the
clusters can be approximated with NFW profiles \citep{navarro1996}:

\begin{equation}
\rho(r)=\frac{\delta_c\rho_c}{\frac{r}{r_s}\left(1+\frac{r}{r_s}\right)^2},
\label{eqn:nfw}
\end{equation}
where $\rho_c=\frac{3H^2(z)}{8\pi G}$ is the critical density of the Universe at redshift $z$, and $r_s$ is the scale radius. We define the radius $r_{200}$ within which the average density of the halo is $200\rho_c$ and the
concentration parameter $c=r_{200}/r_s$. The characteristic overdensity for the halo is

\begin{equation}
\delta_c=\frac{200}{3}\frac{c^3}{\ln(1+c)-c/(1+c)}.
\label{eqn:dc}
\end{equation}
Integrating Eqn.~\ref{eqn:nfw} along the line of sight and defining $x=r/r_s$, an expression for the surface density of the halo can be written\footnote{In detail, this expression assumes that the NFW profile extends to infinity,
although the differences in DM model predictions versus models truncated at the virial radii of the clusters are minimal for our extraction regions.}\citep{wright2000}

\begin{equation}
\Sigma_{NFW}(x)=\left\{
\begin{array}{lr}
\frac{2r_s\delta_c\rho_c}{\left(x^2-1\right)}\left[1-\frac{2}{\sqrt{1-x^2}}\arctanh\sqrt{\frac{1-x}{1+x}}\right] & : x < 1\\
\frac{2r_s\delta_c\rho_c}{3} & : x=1\\
\frac{2r_s\delta_c\rho_c}{\left(x^2-1\right)}\left[1-\frac{2}{\sqrt{x^2-1}}\arctanh\sqrt{\frac{x-1}{x+1}}\right]&:x>1
\end{array}
\right.
\label{eqn:projected}
\end{equation}

Assuming the peak of the NFW profile to coincide with the peak of the X-ray emission from the given cluster, we integrate Eqn.~\ref{eqn:projected} across the \suzaku\ field of view (an $18'\times18'$ square) for the Coma, Virgo and
Ophiuchus clusters and both the core and the confining (an $18'\times18'$ square with a circle of 6~arcmin radius missing from the center for the latter) regions in Perseus. We assume a concentration parameter $c=4.1$
\citep{applegate2014} and the following enclosed virial masses $M_{200}=M\left(r<r_{200}\right)$:

\begin{itemize}
  \item $M_{200}^{\rm Perseus}=6.65\times10^{14}\Msun$ \citep{simionescu2011}
  \item $M_{200}^{\rm Coma}=8.54\times10^{14}\Msun$, calculated from $r_{200}=70\,\text{arcmin}$ \citep{simionescu2013}
  \item $M_{200}^{\rm Virgo}=1.40\times10^{14}\Msun$, calculated from the scaling relations of \citet{arnaud2005} for a mean temperature of 2.3~keV
  \item $M_{200}^{\rm Ophiuchus}=1.47\times10^{15}\Msun$, calculated from the scaling relations of \citet{arnaud2005} for the clusters with $kT>3.5$~keV, for the mean Ophiuchus
  Cluster temperature of 9.4~keV.
\end{itemize}

\begin{table}
\setlength{\extrarowheight}{4pt}
\centering
\caption{Predicted scaling factors for the DM line fluxes for the individual clusters and the extraction regions.  The relative normalization is the ratio of the scaling factors, shown in
the third column, in a given cluster to that in the Perseus core region, which has the most significant detection of the purported line, each divided by the sizes of the respective extraction regions (e.g., a circle with a 6~arcmin
radius for the Perseus core).}
\label{tab:integration}
\begin{tabular}{l|c|cc}
\hline\hline
&$D_L$&$(1+z)M_{\rm proj}/\left(4\pi D_L^2\right)$	&relative norm.\\
&(Mpc)&$\left(10^8M_{\odot}\text{Mpc}^{-2}\right)$&(Perseus core)\\
\hline
\textbf{Perseus} (core)&\multirow{2}{*}{77.7}&$7.21$&\\
\textbf{Perseus} (confining)&&$7.42$&$0.55$\\
\textbf{Coma}&100.7&$13.8$&$0.67$\\
\textbf{Virgo}&15.5&$16.7$&$0.81$\\
\textbf{Ophiuchus}&122.6&$16.3$&$0.79$\\
\hline\hline
\end{tabular}
\end{table}

The results of these calculations are shown in Tab.~\ref{tab:integration}. The systematic uncertainties on these model predictions are probably at the tens of per~cent level. To obtain the relative line normalizations w.r.t. the line
normalization in the Perseus core, we used the ratios of the scaling factors (in the third column of Tab.~\ref{tab:integration}), each divided by the area of the respective extraction regions in the individual clusters (e.g., a circle
with a 6~arcmin radius for the Perseus core). This approach is necessary due to the definition of the normalization of \suzaku\ spectra, which is constant for an object with uniform surface brightness, irrespective of the size of the
extraction region\footnote{For other X-ray observatories, e.g. \cha\ or \xmm, the normalization of such spectrum is proportional to the area of the extraction region.}. Therefore, since the calculations of the scaling factors in the
third column of Tab.~\ref{tab:integration} depend on the extraction region, we need to ``divide out'' this dependence, in order to be able to directly compare the normalizations.

Given these predictions for the relative strengths of emission features due to decaying DM, we tested whether the data for our ensemble of clusters is consistent with such a model. Scaling the strength of putative 3.510~keV (3.592~keV)
emission feature measured in the core (confining) region of Perseus to the Coma, Virgo and Ophiuchus clusters according to the DM model, and fixing the line strengths and energies at these predicted values, we determined new best-fit
C-statistic values for the clusters. These C-statistic values were then compared with those obtained when the line strengths are left free. 

\begin{table}
\setlength{\extrarowheight}{4pt}
\centering
\caption{Change of the fit with a line component, between the case when its normalization has been fixed to the expected value based on the NFW profile according to
Tab.~\ref{tab:integration}, and the case when it was set free. The positions of the features were fixed to their best fit values found in the core of Perseus (3.510~keV,
\emph{Core feature}) and in the confining region of Perseus ($3.592$~keV, \emph{Confining region feature}), respectively. The normalization is given in the units introduced
in~Fig.~\ref{fig:significance}. The shown difference in the C-statistic ($\Delta C$) assumes a perfect knowledge of the scaling of the line strength among the individual
clusters.}
\label{tab:deltacexp}
\begin{tabular}{l|ccc}
\hline\hline
&\multicolumn{2}{c}{line normalization}&$\Delta C$\\
&expected&best-fit&(perfect model)\\
\hline
\multicolumn{4}{c}{\textbf{Core feature}}\\
\hline
\textbf{Perseus} (confining) &$1.99$&$0.26\pm0.21$&65.52\\
\textbf{Coma}&$2.41$&$-0.04\pm0.21$&124.68\\
\textbf{Virgo}&$2.92$&$-0.86\pm0.31$&147.35\\
\textbf{Ophiuchus}&$2.85$&$1.21\pm0.59$&7.86\\
\hline
\textbf{total}&&&345.41\\
\hline
\multicolumn{4}{c}{\textbf{Confining region feature}}\\
\hline
\textbf{Perseus} (core) &$1.01$&$2.30\pm0.43$&9.10\\
\textbf{Coma}&$0.68$&$-0.17\pm0.18$&19.84\\
\textbf{Virgo}&$0.82$&$0.42\pm0.29$&1.91\\
\textbf{Ophiuchus}&$0.80$&$-0.20\pm0.50$&3.97\\
\hline
\textbf{total}&&&34.82\\
\hline\hline
\end{tabular}
\end{table} 

The results of this test are shown in~Tab.~\ref{tab:deltacexp}. We find that
the scaled DM model predictions based on the Perseus core (and confining
region) significantly overestimate the observed line fluxes at these
energies for the other three clusters.  The third column of this table summarizes the difference
in C-statistic between the model with the scaled DM line and with the line
strength free (assuming zero uncertainty in the DM line strength
calculation).  Positive numbers indicate that the fit worsens with the
introduction of the DM line.  Combining the results for Coma, Virgo and
Ophiuchus, and the region in Perseus that the scaling was not based on,
leads to a formal exclusion of the scaled $3.5-3.6$ keV DM line model at
very high significance ($\Delta C>340$ when scaling from the Perseus core,
or $\Delta C=34.82$ when scaling from the confining region).

As an additional test, we examined the internal self-consistency of the Perseus Cluster data, measuring the $\Delta C$ difference between the case when the $\sim$3.55~line properties (the energy and normalization) were fixed at their DM
predicted values, and when both the line energy and normalization were free, i.e. the case described in Sect.~\ref{subs:3.55}. Again, we found a large difference in the C-statistic values, $\Delta C=77.36$, when fitting the confining
region with a DM model scaled from the core region, and $\Delta C=35.41$ when fitting the core region with a DM model scaled from the confining region. Thus the radial variation of the putative DM line appears inconsistent with standard
DM model expectations.

\begin{figure*}
\centering
\begin{minipage}{0.95\columnwidth}
\includegraphics[width=\textwidth]{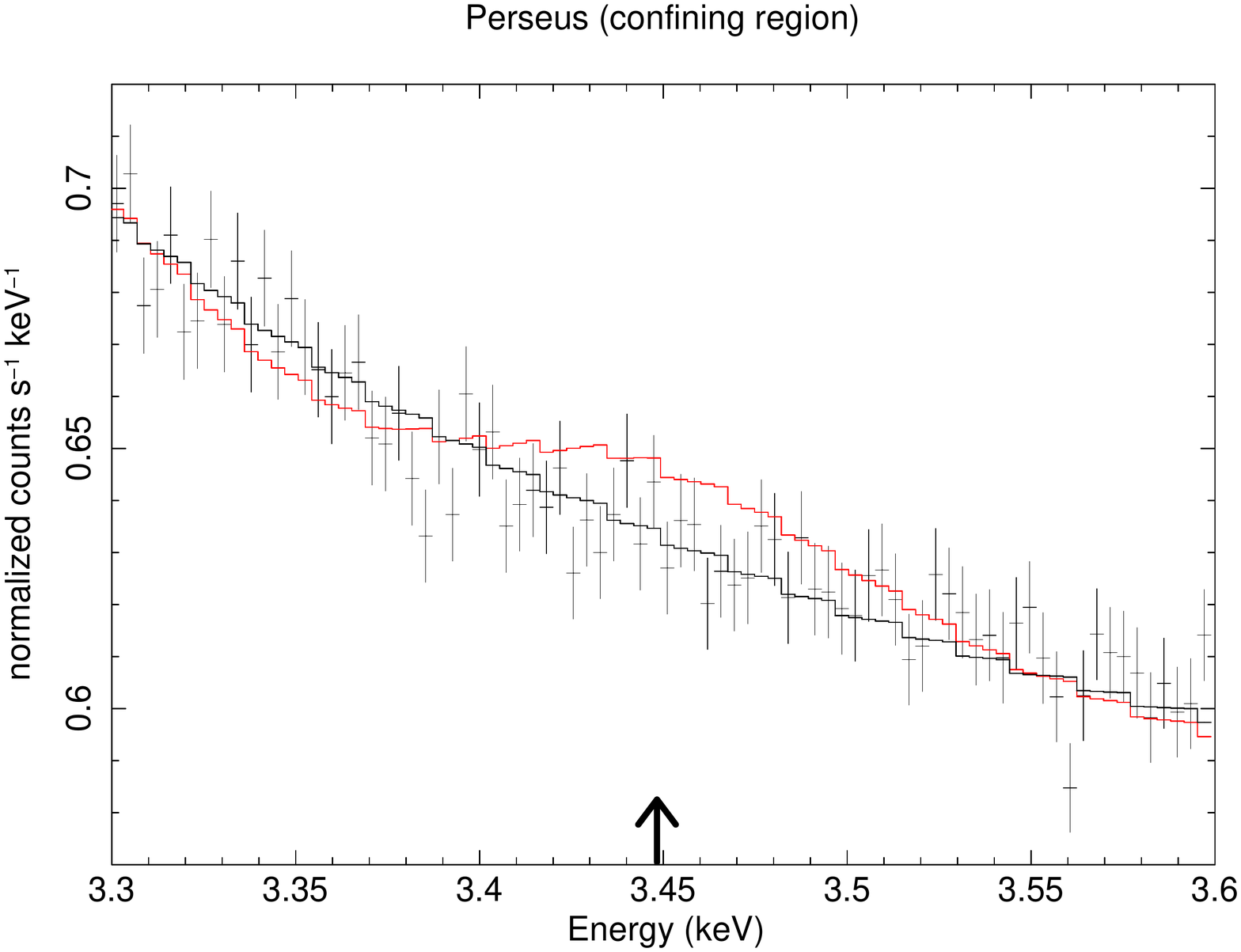}
\end{minipage}
\begin{minipage}{0.95\columnwidth}
\includegraphics[width=\textwidth]{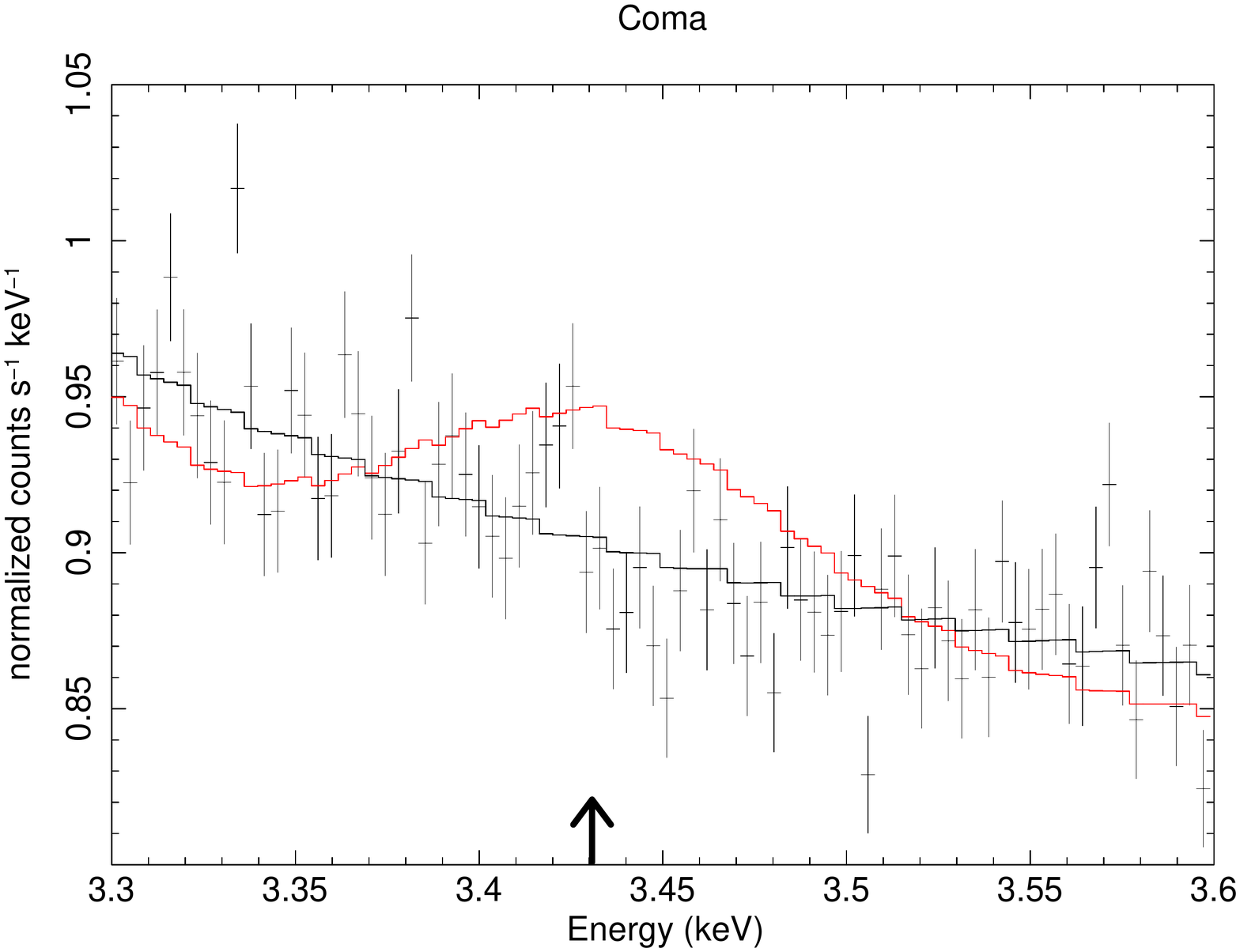}
\end{minipage}
\begin{minipage}{0.95\columnwidth}
\includegraphics[width=\textwidth]{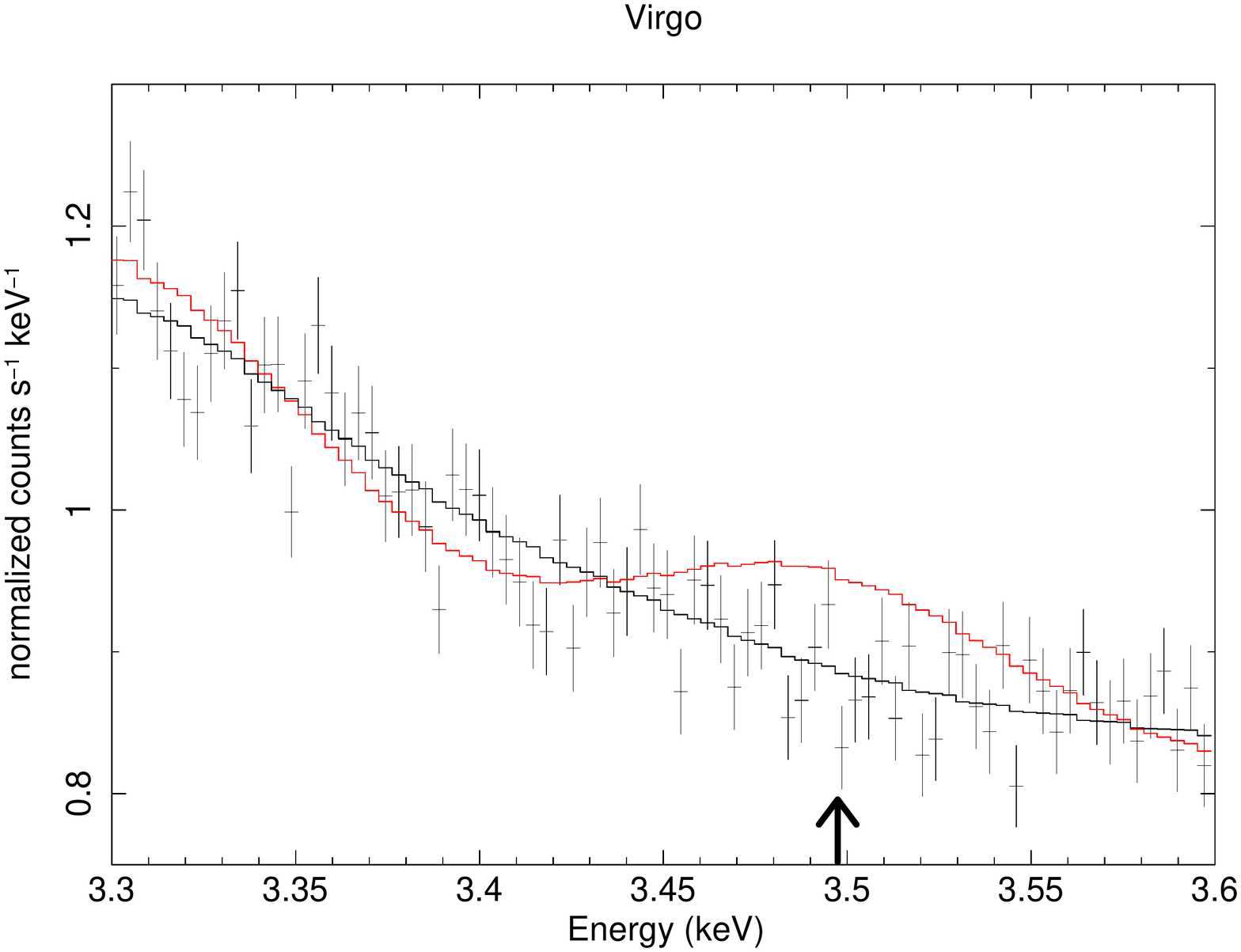}
\end{minipage}
\begin{minipage}{0.95\columnwidth}
\includegraphics[width=\textwidth]{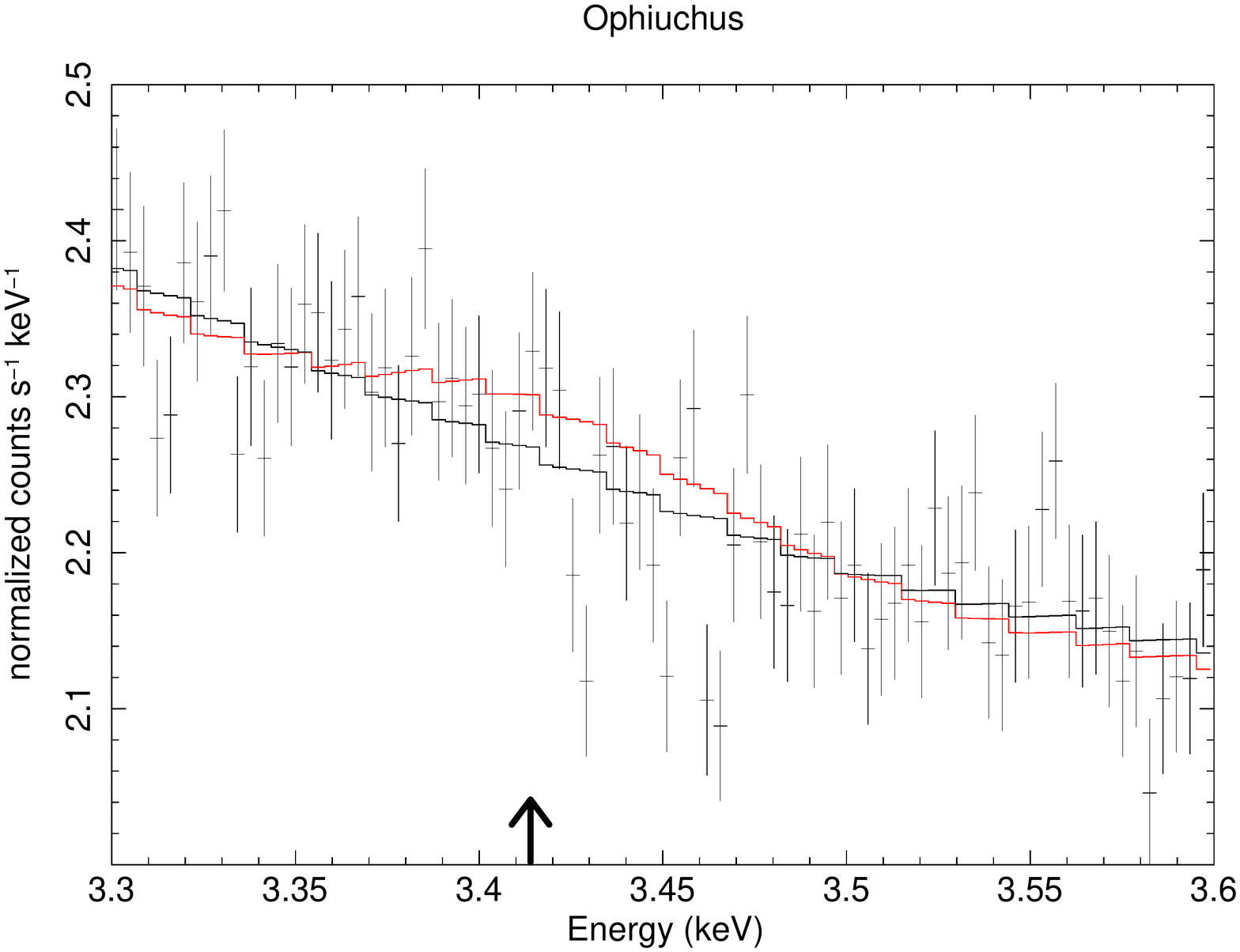}
\end{minipage}
\caption{A detailed view of a narrow part of the cluster spectra in the $\sim$3.5~keV region plotted in the detector frame. The individual panels show the stacked spectra of the individual clusters (\emph{points}), and the best-fit
models including the gaussian line with the normalization set free (\emph{black lines}) and fixed to the predicted value based on the assumption that the line detected in the Perseus core is due to DM decay (\emph{red lines}),
respectively. The black arrows indicates the energy of the gaussian line component in the detector frame. The fitting has been performed in the full $3.2-5.3$~keV energy band. The stacking of the spectra has been done solely for the
illustration purposes and we drew no conclusions based on the analysis of the stacked spectra. The saw-like features seen in the model lines appear due to the limited energy resolution of the response matrices (16eV) and the resulting
discontinuities in the effective area.}
\label{fig:money}
\end{figure*}

We illustrate the tension between the Perseus-scaled decaying DM model  predictions and the Coma, Virgo and Ophiuchus Cluster data in~Fig.~\ref{fig:money}. Here we have stacked the spectra\footnote{The stacking was performed using the
\textsc{addspec} tool.} for the Perseus confining region and fitted the data for each cluster (over the full $3.2-5.3$~keV band) with the baseline plasma model, also including a gaussian component at $E=3.510$~keV with the line strength
scaled according to the decaying DM model normalized to the Perseus Cluster core. The black lines in the figure show the best-fit model when the normalization of the putative DM line is left free. The red lines show the best fit models
with the 3.510~keV line normalization fixed at the predicted value for the decaying DM model, scaled from the Perseus core. In the latter case the fits are poor.

Given the large discrepancies between the DM model-predicted and measured $\sim$3.55~keV line strengths from all individual clusters in our sample, we conclude, that a DM~decay interpretation for the origin of the $\sim$3.55~keV
emission line can be ruled out.

\subsubsection{Other potential origins for the 3.55keV feature: relaxing the
aboundances of other elements}
\label{subs:7abund}

\begin{figure}
\centering
\includegraphics[width=\columnwidth]{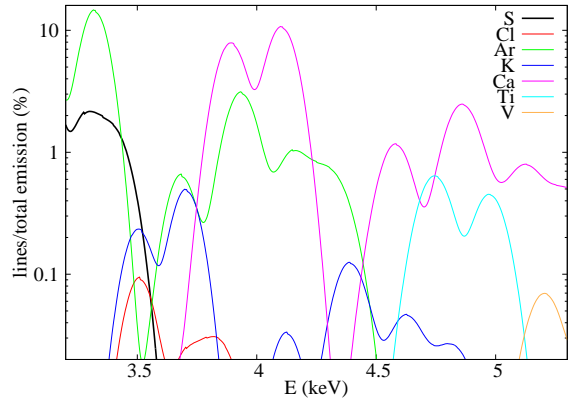}
\caption{Relative contributions of the lines of the individual elements to the total emission, widened by the \suzaku\ response, for all seven elements with tabulated lines in the fitting band $3.2-5.3\,\text{keV}$. The abundances have been set to the Solar values. There are many regions where the lines overlap and, therefore, can influence each other.}
\label{fig:contrib}
\end{figure}

The \suzaku\ spectra for the Perseus Cluster contain
$\sim1.3\times10^7$~cumulative photons (about 85~per~cent of the photons
come from the core region) and are in principle deep enough to allow us to
formally detect emission lines of elements other than S, Ar and Ca.  This is
not possible with any other data set in our sample.  According to
\textsc{AtomDB}, lines of Cl, K, Ti and V with emissivities lower than
$5\times10^{-19}\,\text{photons}\,\text{cm}^3\text{s}^{-1}$ are also present
in our fitting band.  (In detail, a total of 151~line emissivities are
tabulated in our fitting band.) However, not all these lines are included in
our plasma model, since the emissivities of some (predominantly weak) lines
are tabulated for only a narrow range of plasma temperatures and the fitting
code treats their emissivities as zero outside this range.  For example,
only three out of six Cl~lines listed in \textsc{AtomDB} have nonzero model
emissivities at our best-fit temperatures (the remaining three lines are
expected to contribute even less to the total X-ray flux).  To illustrate
the influence of the emission lines on the modeling, we calculated the
fractional contribution of the lines of the individual elements to the
emission across our fitting band, assuming a plasma temperature of 4.8~keV,
Solar abundances of the individual elements, and using the \suzaku\ response
files to account for the line widening.  The results are shown
in~Fig.~\ref{fig:contrib}.

We performed additional fits to both extraction regions of Perseus with the
abundances of the seven elements noted above set free.  In the first case we
included no additional (putative DM) Gaussian component in these fits.  We
found a significant improvement of $\Delta C=-134.13$ and $\Delta C=-35.91$
for four additional degrees of freedom for the core and confining regions,
respectively, with respect to the baseline plasma model.  The measured
values of all free abundances are shown in Tab.~\ref{tab:7abund}.

\begin{table}
\setlength{\extrarowheight}{2pt}
\centering
\caption{Measured abundances of all elements in the Perseus Cluster, that have lines in our fitting band (c.f. Tab~\ref{tab:results}, for the results of the modeling with only the
elements with the strong lines set free). No additional Gaussian line was added to the model. All the quoted errors are statistical only.}
\label{tab:7abund}
\begin{tabular}{c|cc}
\hline\hline
&\textbf{core}&\textbf{confining region}\\
\hline
S &$0.67\pm0.26$&$1.28\pm0.73$\\
Cl&$15.2_{-2.2}^{+2.3}$&$7.00_{-5.56}^{+5.54}$\\
Ar&$0.67\pm0.04$&$0.43\pm0.12$\\
K &$0.95\pm0.44$&$0.24_{-0.72}^{+1.17}$\\
Ca&$0.68\pm0.02$&$0.45\pm0.05$\\
Ti&$2.06\pm0.26$&$2.02_{-0.89}^{+0.90}$\\
V &$12.3_{-1.6}^{+1.5}$&$35.0\pm5.3$\\
\hline\hline
\end{tabular}
\end{table}

The inferred abundances of S and Ar in the Perseus core for the model with all 7~element abundances free are significantly different than those inferred from the previous baseline analysis. This is a consequence  of the complicated
influence that the elemental lines have on each other due to their overlapping energies (given the limited resolution of the \suzaku\ detectors of $\sim$150~eV). This will be discussed further in~Sect.~\ref{subs:disc_lines}. The
abundance of S decreases with the more complex model to approximately solar (or slightly subsolar) values, from the supersolar values of, respectively, 2.12 and 2.24, measured in the baseline analysis. The abundances of all elements
with strong emission lines (S, Ar, Ca) are, in the case of these 7~element fits, broadly consistent with previously reported values to the full-band spectrum \citep[e.g.,][]{tamura2009}.

The results of our 7~element fits imply very high supersolar abundances of Cl and V. As is discussed below, and in more detail in Sect.~\ref{sect:discus}, these very high abundances could in principle be artifacts caused by the
multiphase structure of the gas in and around the Perseus core, as well as modeling degeneracies between the individual elemental lines (see Fig.~\ref{fig:contrib}) and, possibly, systematic errors in the tabulated emissivities.

\begin{figure}
\centering
\includegraphics[width=.95\columnwidth]{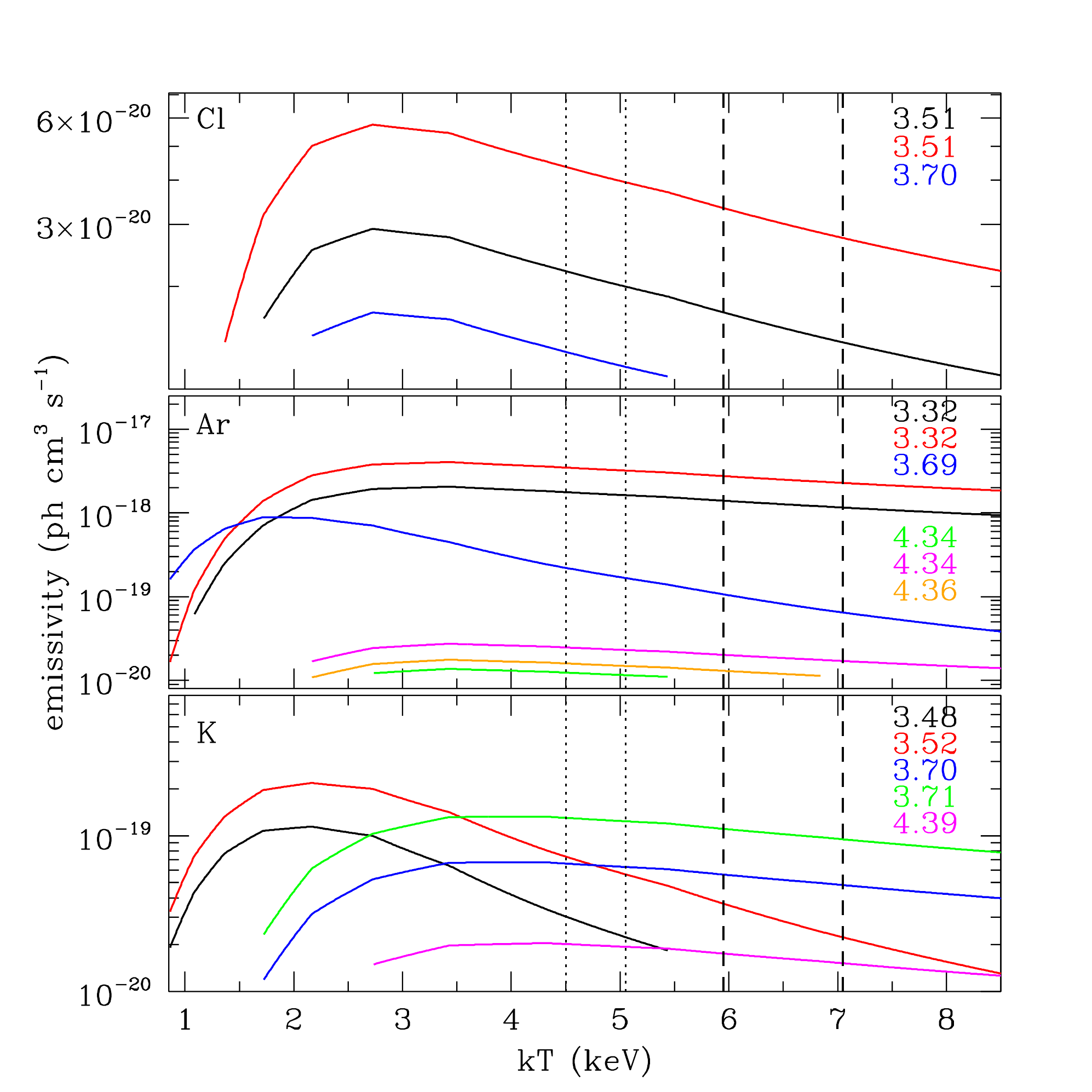}
\caption{Emissivities for the lines of Cl (\emph{top}), Ar (\emph{middle}) and K (\emph{bottom}) as a function of the plasma temperature. The individual line energies in keV are
marked in each panel. The approximate ranges for the best-fit temperatures are marked for the Perseus core (\emph{dotted lines}) and the confining (\emph{dashed lines}) regions.
We show the complete set of the Cl and K lines that appear in our fits, but only those emissivities of the Ar lines that are relevant to the discussion (see text).}
\label{fig:cl_ar}
\end{figure}

Three (two) of the six Cl~lines included in \textsc{AtomDB} have nonzero emissivities at the best-fit temperature of the core (confining) region. These are shown in the top panel of~Fig.~\ref{fig:cl_ar}. Two Cl~\textsc{xvii} lines at
3.51~keV appear in both extraction regions, while the 3.70~keV line of the same ion appears only in the core region. The 3.51~keV line complex is of particular interest. If there is an unidentified emission line present nearby, it could
blend with this complex and cause an upward bias in the Cl~abundance. As a test, we removed the $3.4-3.6$~keV band from the Perseus core spectra and refitted the data\footnote{It is not possible to perform this test with the data from
the confining region, as it would remove all the Cl lines.}. We obtained a marginally higher Cl~abundance of $21.7\pm4.3$ compared to the original case, with no evidence for extra emission around $\sim$3.55~keV. Performing an analogous
test, while removing the $3.6-3.8$~keV band, which contains the 3.70~keV Cl~line, resulted in a formal value of $25_{-3}^{+2}$ for the Cl~abundance. These tests show the complicated way, in which the elemental lines are influenced by
the multitemperature structure of the gas and also by the lines of other elements in the spectrum. This will be discussed in~Sect.~\ref{subs:disc_lines}.

The only V~line in our spectra lies at 5.21~keV, making the measured V~abundance very sensitive to systematic effects, including those stemming from improper continuum fitting. This is supported by the fact, that, in the fit with the
3.70~keV band excluded, as described above, we observed a small increase in the average temperature to $\sim$4.95~keV and a decrease in V~abundance to 9.8. This suggests the presence of extra emission at the high energy end of the
fitting band, with respect to the model, which is modeled by a strong V~line. Possible systematic effects responsible for this surplus emission are discussed in~Sect.~\ref{subs:disc_lines}.

The formal abundances of the other two elements with weak lines, K and Ti, do not immediately look unrealistic (see Tab.~\ref{tab:7abund}). However, they, too, may be subject to bias. Although our results show, that formal measurements
of the chemical abundances of a large set of elements are possible with current data, systematic effects make the interpretation of these results challenging.

\begin{figure*}
\begin{minipage}{0.95\columnwidth}
\includegraphics[width=\textwidth]{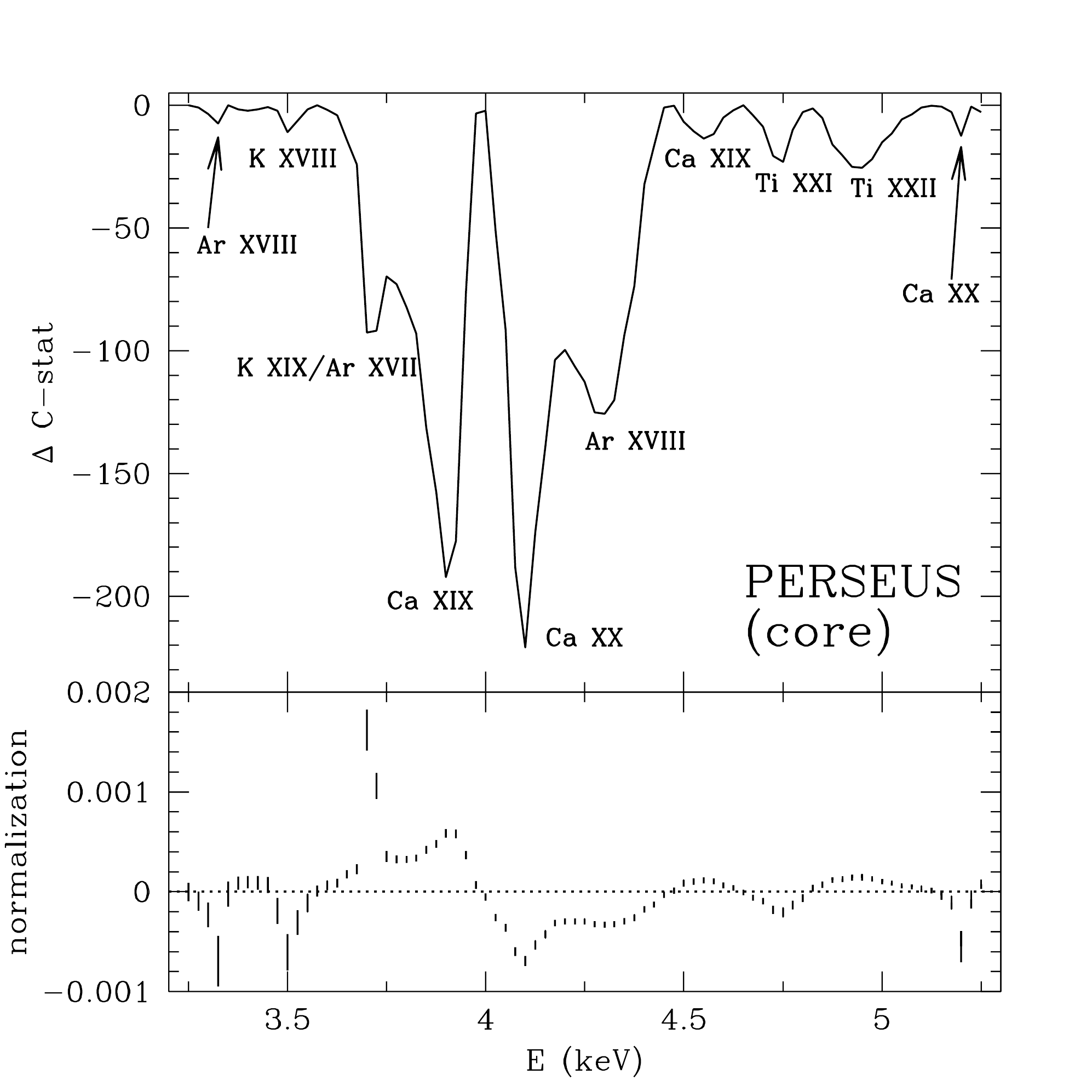}
\end{minipage}
\begin{minipage}{0.95\columnwidth}
\includegraphics[width=\textwidth]{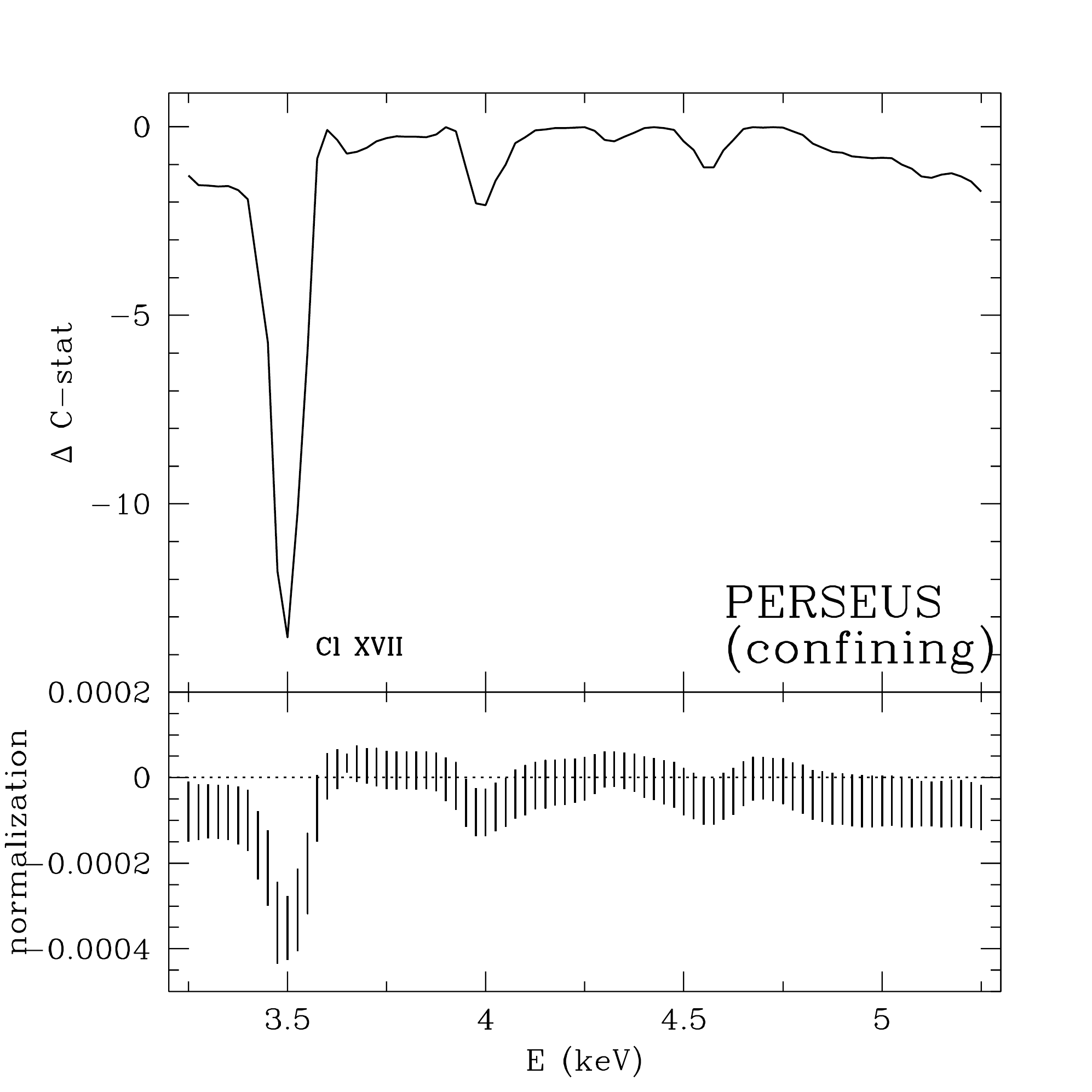}
\end{minipage}
\caption{Improvement of the fit as a function of the rest frame line energy for the Perseus core (\emph{left}) and the confining region (\emph{right}), with seven elemental
abundances (S, Cl, Ar, K, Ca, Ti, V) kept free. For the details of the labels see Sect.~\ref{subs:disc_lines}.}
\label{fig:7abund}
\end{figure*}

We next performed a test for the presence of an additional line, analogous to that introduced in~Sect.~\ref{subs:significance}, but now with the abundances of all seven elements free. The results are shown in Fig.~\ref{fig:7abund} with
the main minima labeled and discussed in~Sect.~\ref{subs:disc_lines}. The curves for the Perseus core in Fig.~\ref{fig:7abund} and the left panel of Fig.~\ref{fig:significance} are broadly similar, except for the near disappearance of
the minima at 3.525~keV and 5.225~keV and the appearance of a new minimum at 3.7~keV in Fig.~\ref{fig:7abund}.

The minimum previously at 3.525~keV has been reduced in depth and shifted by 25~eV, so that it now lies at 3.5~keV, and is now associated with an \emph{absorption} feature. Given, that this change has been achieved simply by including
the abundances of chemical elements with weak emission lines as additional free parameters in the modeling, an elemental origin of the feature seen in~Fig.~\ref{fig:significance}, as opposed to a DM emission line, cannot be excluded.
Other possible reasons for the remaining presence of this minimum are discussed in~Sect.~\ref{subs:disc_lines}.

The curve for the confining region in Perseus is changed dramatically from that in~Fig.~\ref{fig:significance}, and now shows only one, previously unseen, significant feature at 3.5~keV. The best-fit Gaussian at this energy again has a
\emph{negative} normalization and is accompanied by a large increase of the Cl~abundance to a value of $\sim$100. This effect is probably an artifact of the fitting, where the unrealistically strong lines of Cl in the model, caused by
the formally high value of the abundance, are compensated for by the negative residuals from the normalization of the Gaussian line component.

\section{Discussion}
\label{sect:discus}

In the last section we presented evidence suggesting that a previously
reported emission line around 3.55~keV in the Perseus Cluster probably does
not have a dark matter origin, but can instead be explained by elemental
lines, given the current systematic errors in their tabulated intensities. 
In the following section we compare our results with those reported
previously in the literature, and present further discussion of the DM
hypothesis.

\subsection{Arguments relating to the DM Interpretation of Line Emission Around 3.55~keV}
\label{subs:disc_3.55}

In Sect.~\ref{subs:3.55} we saw that, when we took into account only the chemical elements with the line emissivities higher than $5\times10^{-19}\,\text{photons}\,\text{cm}^3\text{s}^{-1}$, we found evidence for emission lines in the
core and the confining regions of the Perseus Cluster with respective best-fit energies of $3.510_{-0.008}^{+0.023}$~keV and $3.592_{-0.024}^{+0.021}$~keV and fluxes of
$2.87_{-0.38}^{+0.33}\times10^{-7}\text{ph}\,\text{s}^{-1}\text{cm}^{-2}\text{arcmin}^{-2}$ and $4.78_{-1.41}^{+1.65}\times10^{-8}\text{ph}\,\text{s}^{-1}\text{cm}^{-2}\text{arcmin}^{-2}$, respectively. Although the $82\pm1$~eV
difference between these line energies is formally significant at the $\sim3\sigma$ level, it is smaller than the resolution of a \suzaku\ detector.

Both of these lines  lie near the feature reported by Bu14, who found an emission line in Perseus, assuming a line energy of 3.57~keV (fixed), with a flux of
$1.22_{-0.36}^{+0.56}\times10^{-7}\text{ph}\,\text{s}^{-1}\text{cm}^{-2}\text{arcmin}^{-2}$ (using a circular extraction region with the radius of 700~arcsec). Bo14 found an emission line at $3.518_{-0.022}^{+0.019}$~keV in a combined
analysis of M31 and the Perseus Cluster. However, the latter work quotes no value for the line flux.

The energies of the emission features we detect in the Perseus Cluster are
formally inconsistent with the values of Bu14 and Bo14 (with the exception
of the feature we found in the confining region, whose energy is consistent
with the value found by~Bo14).  However, the difference is less than the
energy resolution of the XIS detectors onboard \suzaku\ ($\sim$150~eV).

At face value, the fluxes of the emission features reported here and in Bu14
are also broadly consistent.  The differences in the results are smaller
than the systematic uncertainties in the line flux reported by~Bu14 (their
Tab.~5), where the fluxes of the 3.57~keV feature measured by the different
\xmm\ detectors differ up to a factor of a few.  Moreover, the differences
can in part be caused by the different energies in the reported lines.  For
example, when we fixed the line energy in the Perseus core at 3.57~keV (the
energy used in~Bu14), our measured line flux decreases by $\sim$30~per~cent,
to
$\left(2.08\pm0.34\right)\times10^{-7}\text{ph}\,\text{s}^{-1}\text{cm}^{-2}\text{arcmin}^{-2}$. 
Finally, Bu14~used different spectral extraction region from the analysis
presented here, which can also introduce differences between the measured
fluxes.

The arguments discussed above suggest a common origin of the $\sim$3.55~keV
lines measured here in both extraction regions of the Perseus Cluster and
those reported in Bu14 and Bo14.  In this case, the line strength scaling
arguments presented in~Sect.~\ref{subs:expectation} challenge the DM
interpretation for the origin of the line.  Specifically, we showed, that
the line strength does not follow the expected scaling relations based on
NFW modeling of the DM profiles in the Coma, Virgo and Ophiuchus clusters or
indeed radially in Perseus itself.  The assumption of DM origin for the
3.510~keV line, with a line strength scaled to the value measured in the
Perseus Cluster core leads to an expected strength of this line that
over-predicts the signal in the other clusters.  The upper limits of the
measured line strengths in each cluster are inconsistent with the expected
line strength at more than 99.5~per cent level.  The Virgo Cluster, where we
observe a significantly negative normalization of the line at 3.510~keV (see
Tab.~\ref{tab:deltacexp}) is particularly challenging to the dark matter
interpretation.  The discrepancy between the expected and measured line
strengths is less pronounced but still significant when normalizing the DM
model predictions to the Perseus confining region.

Formally, the tension between the expected and measured line strength for the two Perseus Cluster regions also strongly disfavours the DM origin interpretation. In~Sect.~\ref{subs:expectation} we showed, that we should expect the line
strength from the Perseus confining region to be approximately half that in the Perseus core. However, we observe a putative DM line flux in the Perseus core that is up to~$\sim$14 times higher than the flux from the confining region,
at the same energy. In fact, due to the point-spread function (PSF) contaminating the confining region spectra with core emission to some degree (which increases the expected surface brightness of the confining region), the actual ratio
between the fluxes is probably higher, increasing the discrepancy.

Finally, including a larger set of heavy element lines in the modeling significantly reduces the requirements for the $\sim$3.55~keV line in Perseus.

\subsection{Systematic Effects Influencing the Detailed Chemical Composition of the Perseus Cluster ICM}
\label{subs:chemistry}

The Perseus Cluster data allow us to probe the detailed chemical composition
of the ICM.  \textsc{AtomDB} contains 151~tabulated lines in the fitting
band of $3.2-5.3$~keV for a total of seven chemical elements: S, Cl, Ar, K,
Ca, Ti and V.  We were able to formally constrain nonzero abundances for all
of these elements, as shown in~Tab.~\ref{tab:7abund}.  However, as discussed
in~Sect.~\ref{subs:7abund}, the formal values of the abundances, especially
for the elements with weak emission lines -- Cl, K, Ti and V -- are probably
dominated by systematic effects.  In this section we discuss potential
origins for these effects in detail, including the presence of multiphase
gas, overlapping line energies, systematic uncertainties in the tabulated
line emissivities and artifacts of the fitting process itself.

We expect the multitemperature structure of the ICM in the Perseus core
(that we cannot account for with our single temperature model) to introduce
an upward bias in the measured abundances of all elements.  The emissivities
of all emission lines, except for the lines of~Ti~\textsc{xxii} (at 4.97~keV
and 4.98~keV) and the weak lines of Ca~\textsc{xx} with energies higher than
4.86~keV, peak at temperatures lower than the average best-fit temperature
of $\sim$4.8~keV that we measure in the Perseus core.  Therefore, these
lines will appear overabundant in this region, where we know cooler gas is
present \citep{sanders2007}.

Due to the limited energy resolution of the \suzaku\ CCD detectors
($\sim$150~eV) and a relatively high number of the elemental lines in our
fitting band, we observe a number of overlaps between the individual lines,
which can be seen in~Fig.~\ref{fig:contrib}.  Combined with the systematic
uncertainty in the tabulated emissivities of the weak, and especially the
satellite lines, which can reach $\sim10-20$~per~cent, these can introduce
bias to the measured abundances.

We can examine the very high formal values of the Cl~abundance of $15\pm2$,
reported in~Sect.~\ref{subs:7abund}, as a potential example of these
effects.  In the Perseus core, Cl~lines are present at 3.51~keV and
3.70~keV, and at their respective energies they contribute $\sim$2\% and
$\sim$0.5\% to the total spectral signal, assuming the measured abundance of
Cl (cf.  Fig.~\ref{fig:contrib} for the line contribution to the total
signal at the Solar abundances).  In~Sect.~\ref{subs:7abund} we showed, that
the high Cl~abundance is formally robust and not caused by a presence of an
unidentified emission line blending with one of the Cl~lines.  (In detail,
we showed that if we remove either of the energy bands containing the
Cl~lines ($3.4-3.6$~keV and $3.6-3.8$~keV for the $3.51$~keV and the
$3.70$~keV lines, respectively), the abundance does not decrease.)

The lines of three elements, Cl, Ar and K, contribute to the spectrum in the
$\sim$3.70~keV band.  The abundance of Ar~is predominantly determined from
the 3.32~keV lines, which constitute the strongest line complex in our
spectra.  Their emissivities peak at the temperature of $\sim$3.4~keV, while
the emissivities of the Ar~lines in the $\sim$3.70~keV band peak at lower
temperatures of $<1.7$~keV.  In the presence of relatively cool gas in the
Perseus core, that our single temperature model cannot account for, the
latter set of Ar~lines would be underpredicted, causing surplus emission in
the $\sim$3.70~keV energy band.  [Alternatively, if there are differences
between the true and the tabulated values of the emissivities of the
Ar~lines in the $\sim$3.70~keV region, specifically, if the tabulated values
underestimate the true value, a surplus emission would appear at the
energies around $\sim$3.70~keV, since the Ar~abundance is well determined
from the strongest lines at 3.32~keV.] The fitting code can mistakenly
interpret the surplus at $\sim$3.70~keV as having been produced by the lines
of extremely abundant Cl.

In the case of the systematic influence on the Cl~lines at 3.51~keV, the
scenario is more complicated.  Qualitatively, the Ar~lines could, in a
similar fashion as discussed above, influence the measured abundances of S
(overlap in the $\sim$3.3~keV band, see Fig.~\ref{fig:contrib}) and K
(overlap in the $\sim$3.7~keV and $\sim$4.3~keV bands).  The lines of the
two latter elements contribute to the emission in the $\sim$3.5~keV region
and could, in turn, have an influence on the apparent Cl~line strength.

We identified extra emission at the high energy end of the spectrum with respect to the model as the reason for the high formal value of V~abundance in~Sect.~\ref{subs:7abund}. This extra emission may appear due to a systematic error in
the calibration of the effective area as a function of energy, which would primarily affect the edge of the fitting band and may subsequently lead to an apparent surplus of emission at the high energy end of our fitting band.
Alternatively, if emission with temperature higher than $\sim$4.8~keV, measured in the Perseus core, was present in our spectra (e.g. in projection from the higher parts of the cluster atmosphere, or from an additional power-law
spectral component), the continuum would fall off slower than predicted by our single temperature model, again leading to surplus emission at the high energy end of the spectrum. Similarly, the emission from the confining region could
be contaminated by the colder emission from the cluster core due to the PSF effects (see Sect.~\ref{subs:disc_3.55}), causing a downward bias in the best fit temperature and a quick fall off of the model spectrum with respect to the
data, which would then lead to the same effect. Unfortunately, the data do not allow us to perform a multitemperature fit to test this possibility in either of these regions.

\subsection{Spectral Signatures of the Ionization Balance and the Multitemperature Structure}
\label{subs:disc_lines}

Figs.~\ref{fig:significance} and \ref{fig:7abund} exhibit multiple minima
for each cluster, showing formally significant improvements to the fit by
adding a gaussian line component at a given energy to the model.  Here we
discuss possible causes for these features.

As shown in~Sect.~\ref{subs:7abund}, some of these minima can be removed, or
reduced in depth, with more sophisticated models for the abundances of
elements with weak spectral lines.  This was seen for the 3.525~keV and
5.225~keV features in the Perseus core and all of the minima in the Perseus
confining region.

However, multiple features remain present in these data even after
accounting for the elements with weak lines.  Since detailed fitting was
only possible with the Perseus data, and given, that the Perseus core
spectra retained many of its features when fitting with three or seven free
elemental abundances, we predominantly focus our discussion on the results
in the left panel of~Fig.~\ref{fig:7abund}.

The left panel of~Fig.~\ref{fig:7abund}, for the Perseus core, is rich in
features.  All the minima lie at, or near, emission lines of various
elements that have been labeled in the figure.  We point out, that for each
labeled element, a pair of ions is present, e.g.  Ca~\textsc{xix} and
Ca~\textsc{xx}.  Moreover, if a given element is associated with multiple
minima, the signs of the best-fit normalization of the gaussian component in
the model are opposite for each ion of the pair, e.g.  the minima at the
position of Ca~\textsc{xix} (\textsc{xx}) lines are always associated with a
positive (negative) line normalization.

This suggests problems in the assumed ionization balance for the elements in the cluster core. If the ratio of different ions of the given element differs from the expected equilibrium value, its measured abundance will lie between the
values that would be determined from the lines of individual ions. The additional line component in our model allows the fit to compensate for this discrepancy. 

In effect, we are underestimating the abundances of Ar~\textsc{xvii} (with a strong line at 3.69~keV), K~\textsc{xix} (strongest line at 3.71~keV), Ca~\textsc{xix} (with a complex of lines around 3.9~keV and another at 4.6~keV) and
Ti~\textsc{xxii} (unresolved lines at 4.97~keV and 4.98~keV), and overestimating the abundances of Ar~\textsc{xviii} (complex of lines at the low energy end of our fitting band with the strongest line at 3.32~keV), K~\textsc{xviii}
(complex of lines around 3.5~keV), Ca~\textsc{xx} (complex of lines around 4.1~keV with the strongest at 4.107~keV) and Ti~\textsc{xxi} (complex of lines around 4.75~keV).

The features in~Fig.~\ref{fig:7abund} could also in part be explained by the multitemperature structure of the projected emission from the Perseus core, that we are unable to model fully. In the multitemperature gas, different gas
phases will contain elemental ions in different concentrations, as governed by the local temperature, and the individual ion populations will contribute to the emission lines with different strengths. Therefore, it may not be possible
to assign a single temperature to the gas based on the line strengths. This was seen by~\citet{jeltema2014}, who used the emission line strength ratios quoted by Bu14 to estimate the plasma temperature in the Perseus observations, and
obtained a range of inconsistent values. However, we note, that \citet{jeltema2014} were using the line strength estimates provided by the \textsc{webguide} tool of \textsc{AtomDB}, that can significantly differ from values obtained
using the full calculation \citep{bulbul2014b}.

The multitemperature structure of the gas in Perseus has been observed
before \citep[e.g.,][]{sanders2007}, which lends support to the above
scenario.  However, a relatively complicated temperature distribution would
be required to explain the multitude of minima in Fig.~\ref{fig:7abund}. 
For instance, lower temperature gas, unaccounted for in the analysis, could
cause the emission lines from ions with a lower stage of ionization to be
present in higher concentration, qualitatively describing the behaviour of
the Ca ions in Fig.~\ref{fig:7abund}.  However, an opposite process seems to
be influencing Ti, where the gaussian component seems to be compensating for
a too high measured abundance of Ti~\textsc{xxi} by fitting to a negative
value and vice versa for Ti~\textsc{xxii}.  If this is also due to a
multi-phase structure of the gas, it would suggest the additional presence
of a third spectral component corresponding to a temperature that is higher
than the measured average.

The example scenarios outlined in Sect.~\ref{subs:chemistry} and this
section illustrate the complicated interaction between various systematic
effects on the spectral modeling.  To attempt to disentangle these effects
is beyond the capabilities of \suzaku, given the limited spectral
resolution.  Observations with next generation detectors, especially the
calorimeters onboard \emph{Astro-H}, will be required to provide more
robust constraints on the chemical composition and thermal structure of the
ICM.

\section{Conclusions}
\label{sect:concl}

In light of the recent claims of the discovery of an unidentified emission
line at $\sim$3.55~keV and its possible dark matter origin, we have used
\suzaku\ observations to search for unaccounted emission lines in the
$3.2-5.3$~keV energy band in the spectra of the four X-ray brightest galaxy
clusters.  Our main conclusions are:

\begin{itemize} \item Using a single temperature model with the abundances
of S, Ar and Ca as free parameters, we detected emission lines at
$E=3.51^{+0.02}_{-0.01}$~keV in the core, and $E=3.59\pm0.02$~keV in the
regions surrounding the core, of the Perseus Cluster.  The properties of
these lines are broadly consistent with the previously reported unidentified
lines in this energy band.  If these lines are produced by the dark matter
particle decay, they should be observed in other clusters with a strength
proportional to the projected mass of the dark matter within our field of
view.  However, scaling from the Perseus Cluster core, we measured the upper
limits of the strength of these lines in the other clusters that are in each
case inconsistent with the DM model predictions at more than 99.5~per cent
level, disfavouring the DM origin hypothesis.  The radial variation of the
putative DM line strength in Perseus is also in apparent tension with the DM
model hypothesis for the 3.55~keV line.

\item Our analysis suggests that the $\sim$3.55~keV line features measured
in the Perseus Cluster may have an elemental origin.  The observations of
the Perseus Cluster allowed us to formally measure the abundances of seven
elements: S, Cl, Ar, K, Ca, Ti and V, with emission lines in the
$3.2-5.3$~keV energy band.  Even though we obtain a significant nonzero
value for the abundance for each of the seven elements, these measurements
are, particularly for the elements with relatively weak emission lines (Cl,
K, Ti and V), heavily influenced by systematic effects, including the
presence of multitemperature structure in the Perseus Cluster; overlaps
between the individual lines due to the limited resolution of the \suzaku\
detectors; systematic errors in the tabulated values of the line
emissivities; uncertainties in the detector effective area calibration; and
problems with the assumed ionization balance.  These systematics are in big
part exacerbated by the limited resolution of the CCD detectors
($\sim$150~eV).  Next generation X-ray detectors with a $\sim$few~eV
resolution, starting with \emph{Astro-H}, will be needed to perform more
robust measurements of the chemical composition of the ICM.

\end{itemize}

\section*{Acknowledgments}

This work was supported in part by  the US Department of Energy under
contract number DE-AC02-76SF00515.  This research has made use of data
obtained from the \suzaku\ satellite, a collaborative mission between the
space agencies of Japan (JAXA) and the USA (NASA).

\bibliographystyle{aa}
\bibliography{clusters}

\appendix

\end{document}